\pdfoutput=1
\documentclass[fleqn,usenatbib,useAMS]{mnras}

\usepackage{graphicx}	
\usepackage{amsmath}	
\usepackage{amssymb}	
\usepackage{multicol}        
\usepackage{bm}		
\usepackage{pdflscape}	
\usepackage{ulem}
\usepackage{inputenc}

\newcommand{\beq}{\begin{equation}}
\newcommand{\beqa}{\begin{eqnarray}}
\newcommand{\eeq}{\end{equation}}
\newcommand{\eeqa}{\end{eqnarray}}

\newcommand{\simgt}{\lower.5ex\hbox{$\; \buildrel > \over \sim \;$}}
\newcommand{\simlt}{\lower.5ex\hbox{$\; \buildrel < \over \sim \;$}}

\newcommand{\bd}[1]{\mbox{\boldmath $#1$}}

\usepackage[T1]{fontenc}
\usepackage{ae,aecompl}

\usepackage{newtxtext,newtxmath}

\newcommand{\rev}[1]{\textcolor{black}{#1}}

\title[Deep learning for HSC lensing map]{
Noise reduction for weak lensing mass mapping: 
An application of generative adversarial networks to Subaru Hyper Suprime-Cam first-year data}

\author[M. Shirasaki et al.]
{Masato Shirasaki$^{1,2}$\thanks{E-mail: masato.shirasaki@nao.ac.jp}
Kana Moriwaki$^{3}$,
Taira Oogi$^{4}$,
Naoki Yoshida$^{3,5,6,7}$,
Shiro Ikeda$^{2,8}$,
\newauthor
Takahiro Nishimichi$^{5,9}$,
\\
$^{1}$National Astronomical Observatory of Japan, 
Mitaka, Tokyo 181-8588, Japan\\
$^{2}$The Institute of Statistical Mathematics, Tachikawa, Tokyo 190-8562, Japan\\
$^{3}$Department of Physics, University of Tokyo, Tokyo 113-0033, Japan\\
$^{4}$Institute of Management and Information Technologies, Chiba University, Chiba 263-8522, Japan\\
$^{5}$Kavli Institute for the Physics and Mathematics of the Universe (WPI), University of Tokyo, Kashiwa, Chiba 277-8583, Japan\\
$^{6}$Institute for Physics of Intelligence, University of Tokyo, Tokyo 113-0033, Japan\\
$^{7}$Research Center for the Early Universe, Faculty of Science, University of Tokyo, Tokyo 113-0033, Japan\\
$^{8}$Department of Statistical Science,
Graduate University for Advanced Studies, 10-3 Midori-cho, Tachikawa, Tokyo 190-8562, Japan\\
$^{9}$Center for Gravitational Physics, Yukawa Institute for Theoretical Physics, Kyoto University, Kyoto 606-8502, Japan
}

\date{}

\pubyear{2021}

\begin{document}
\label{firstpage}
\pagerange{\pageref{firstpage}--\pageref{lastpage}}
\maketitle

\begin{abstract}
We propose a deep-learning approach based on 
generative adversarial networks (GANs) to reduce noise in weak lensing mass maps under realistic conditions. 
We apply image-to-image translation using conditional GANs to the mass map obtained from the first-year data of Subaru Hyper Suprime-Cam (HSC) survey. 
We train the conditional GANs by using 25000 mock HSC catalogues that directly incorporate a variety of observational effects. 
We study the non-Gaussian information in denoised maps using
one-point probability distribution functions (PDFs) and also perform matching analysis for positive peaks and massive clusters.
An ensemble learning technique with our GANs is successfully applied to reproduce the PDFs of the lensing convergence. About $60\%$ of the peaks in the denoised maps with height greater than $5\sigma$ have counterparts of massive clusters within a separation of 6 arcmin.
We show that PDFs in the denoised maps are not compromised by 
details of multiplicative biases and photometric redshift distributions, 
nor by shape measurement errors, 
and that the PDFs show stronger cosmological dependence compared to the noisy counterpart.
We apply our denoising method to a part of the first-year HSC data to show that the observed mass distribution is statistically consistent with the prediction from the standard $\Lambda$CDM model.
\end{abstract}





\section{Introduction}

Impressive progress has been seen in observational cosmology in the past decades.
An array of multi-wavelength astronomical data 
have established the standard model of our universe, referred to as $\Lambda$CDM model, with precise determination of major cosmological parameters.
The nature of the main energy contents in our universe remains unknown, however. Invisible mass component called dark matter is needed to 
explain the formation of large-scale structures in the universe \citep{Clowe:2003tk, Ade:2015xua}, 
and an exotic form of energy appears to be responsible for 
the accelerating expansion of the present-day universe \citep{Huterer:2017buf}.

With an important aim at revealing the nature of dark matter and the late-time cosmic acceleration, 
a number of astronomical surveys are ongoing and planned.
Accurate measurement of cosmic lensing shear signals is  one of the primary goals of such galaxy surveys including
the Kilo-Degree Survey (KiDS\footnote{\url{http://kids.strw.leidenuniv.nl/index.php}}), 
the Dark Energy Survey (DES\footnote{\url{https://www.darkenergysurvey.org/}}),
and the Subaru Hyper Suprime-Cam Survey (HSC\footnote{\url{http://hsc.mtk.nao.ac.jp/ssp/}}),
even for upcoming projects including
the Nancy Grace Roman Space Telescope
(Roman\footnote{\url{https://roman.gsfc.nasa.gov/}}),
the 
Legacy Survey of Space and Time on Vera C. Rubin Observatory
(LSST\footnote{\url{https://www.lsst.org/}}),
and Euclid\footnote{\url{http://sci.esa.int/euclid/}}.
 
The large-scale matter distribution in the universe can be reconstructed 
through measurements of lensing shear signals by collecting 
and analysing a large set of galaxy images \citep{Tyson:1990yt, Kaiser:1992ps, Schneider:1996ug}. 
Although the image distortion of individual galaxies is typically very small, 
it is possible 
to infer the distribution of underlying matter density in an unbiased way 
by averaging over many galaxies.
However, there are well-known challenges in practice when extracting rich cosmological information from the reconstructed matter distribution.
Non-linear gravitational growth of the large-scale structure 
renders the statistical properties of the weak lensing signal complicated. 
Numerical simulations have shown that 
popular and powerful statistics for random Gaussian fields 
such as power spectrum are not able to fully describe 
the cosmological information imprinted in weak lensing maps \citep{Jain:1999ir, Hamana:2001vz, Sato:2009ct}.
To extract and utilise the so-called non-Gaussian information, various approaches have been proposed \citep[e.g.][]{Matsubara:2000dg, Sato:2001cb, Zaldarriaga:2002qt, Takada:2002hh, Pen:2003vw, Jarvis:2003wq, Wang:2008hi, 2010MNRAS.402.1049D, 2010PhRvD..81d3519K, 2010ApJ...719.1408F, Shirasaki:2013zpa, 2015A&A...576A..24L, 2015PhRvD..91j3511P, 2018arXiv181002374C, 2017arXiv170705167S, Gupta:2018eev, 2019NatAs...3...93R},
but no single statistic can capture the full information, unfortunately.

In practice, an observed weak lensing map is contaminated 
with noise arising from intrinsic galaxy properties and observational conditions.
The former is commonly called as shape noise, which 
compromises the original physical effect.
It is known that the noise effect can be robustly estimated and 
can also be mitigated for a Gaussian field \citep{Hu:2000ax, Schneider:2002jd}, 
but little is known about the overall
impact of the shape noise on a non-Gaussian field.
Non-Gaussian information can potentially be a powerful probe to 
test the $\Lambda$CDM model
and variant cosmological models even in the presence of shape noise \citep{Shirasaki:2016twn, Liu:2018dsw, Marques:2018ctl}.
For instance, \citet{2020arXiv200706529Z} show that cosmological inference based on convolutional
neural networks relies on the information carried by high-density regions where the noise is less important.
Clearly, it is important to devise a noise reduction method in order
to maximise the science return from ongoing and future wide-field lensing surveys.

A straightforward way of mitigating the shape noise is 
to smooth a weak lensing map over a large angular scale (e.g. $\sim20-30 \, {\rm arcmins}$ in \citet{Vikram:2015leg, Chang:2017kmv}), 
but the smoothing itself also erases the non-Gaussian information in the map \citep{Jain:1999ir, Taruya:2002vy}.
A novel approach has been proposed to keep a high angular resolution
of $\sim1 {\rm arcmin}$ while preserving non-Gaussian information \citep{Shirasaki:2018thk}.
The method is based on a deep-learning framework called 
conditional generative adversarial networks (GANs) \citep{2016arXiv161107004I}.
Thanks to the expressive power of deep neural networks, 
conditional GANs can denoise a weak lensing map on a pixel-by-pixel basis (see also \citet{Jeffrey:2019fag, 2020arXiv201108271R} for similar study with a deep learning method).
In exchange for its versatility, deep learning methods need to 
be validated thoroughly before applied to real observational data.
However, validation of deep learning for lensing analyses has not been fully explored so far.
To examine and improve the capability of denoising with
deep learning,
we need to study the statistical properties of denoised weak lensing maps using conditional GANs.

In this paper, we construct and test conditional GANs and  
apply to the real galaxy imaging data from Subaru HSC
survey \citep{Aihara:2017paw}. We use a large set of realistic mock HSC catalogues \citep{Shirasaki:2019gya} to train the GANs.
We test the denoising method using 1000 test data and assess possible systematic errors in the denoising process.
We investigate non-Gaussian information in the denoised maps 
by using realistic simulations of gravitational lensing.
For the first time, we evaluate generalisation errors in the denoising process by varying several characteristics in the mock HSC catalogues.
After stress-testing, we apply our GANs to the real HSC data and
study the cosmological implication of the reconstructed 
large-scale matter distribution.

The rest of the present paper is
organised as follows. 
In Section~\ref{sec:WL}, we summarise the basics of gravitational lensing. 
Section~\ref{sec:data} describes the HSC data as well as our numerical simulations used for training and testing GANs.
In Section~\ref{sec:analysis}, we explain the details of our training strategy of GANs.
Before applying the GANs to the real HSC data, we perform thorough tests. We present the results in Section~\ref{sec:prop_denoise_map}.
In Section~\ref{sec:results}, we show the denoised map for the HSC data.
Concluding remarks and discussions are given in Section~\ref{sec:con}. 

\section{Weak gravitational lensing}
\label{sec:WL}

\subsection{Basics}

We first summarise the basics of gravitational lensing 
induced by the large-scale structure.
Weak gravitational lensing effect is characterised by
the distortion of the image of a source object, formulated
by the following 2D matrix:
\beq
A_{ij} = \frac{\partial \beta^{i}}{\partial \theta^{j}}
           \equiv \left(
\begin{array}{cc}
1-\kappa -\gamma_{1} & -\gamma_{2}  \\
-\gamma_{2} & 1-\kappa+\gamma_{1} \\
\end{array}
\right), \label{eq:distortion_tensor}
\eeq
where $\bd{\theta}$ represents the observed position of a source object,
$\bd{\beta}$ is the true position, 
$\kappa$ is the convergence, and $\gamma$ is the shear.
In the weak lensing regime (i.e., $\kappa, \gamma \ll 1$), 
each component of $A_{ij}$ can be related to
the second derivative of the gravitational potential $\Phi$
\citep{2001PhR...340..291B}.
Using the Poisson equation and the Born approximation, 
one can express the weak lensing convergence field as the weighted 
integral of matter over-density field $\delta_{\rm m}(\bd{x})$:
\beq
\kappa(\bd{\theta})
= \int_{0}^{\chi_{H}} {\rm d}\chi \ q(\chi)\delta_{\rm m}(\chi,r(\chi)\bd{\theta}), \label{eq:kappa_delta}
\eeq
where $\chi$ is the comoving distance, 
$\chi_{H}$ is the comoving distance up to $z$,
and $q(\chi)$ is called lensing kernel.
For a given redshift distribution of source galaxies,
the lensing kernel is expressed as
\beq
q(\chi) = \frac{3}{2}\left( \frac{H_{0}}{c}\right)^2 \Omega_{\rm m0} \frac{r(\chi)}{a(\chi)}\, \int_{\chi}^{\chi_{H}} {\rm d}\chi^{\prime} p(\chi^{\prime})\frac{r(\chi^{\prime}-\chi)}{r(\chi^{\prime})}, \label{eq:lens_kernel}
\eeq
where $r(\chi)$ is the angular diameter distance
and $p(\chi)$ represents the redshift distribution of source galaxies
normalised to $\int_{0}^{\chi_H} {\rm d}\chi \, p(\chi) =1$.

\subsection{Smoothed lensing convergence map}
\label{subsec:reconst_kappa}

In optical imaging surveys, galaxies' shapes (ellipticities) are commonly used to estimate the shear component $\gamma$ in Eq.~(\ref{eq:distortion_tensor}).
Since each component in the tensor $A_{ij}$
is given by the second derivative of the gravitational potential, 
one can reconstruct the convergence field 
from the observed shear, in Fourier space, as 
\beqa
\hat{\kappa}(\bd{\ell})
=\frac{\ell_{1}^2-\ell_{2}^2}{\ell_{1}^2+\ell_{2}^{2}}\hat{\gamma}_{1}(\bd{\ell})
+\frac{2\ell_{1}\ell_{2}}{\ell_{1}^2+\ell_{2}^{2}}
\hat{\gamma}_{2}(\bd{\ell}),
\label{eq:gamma2kappa}
\eeqa
where $\hat{\kappa}$ and $\hat{\gamma}$
are the convergence and shear in Fourier space, 
and $\bd{\ell}$ is the wave vector with components $\ell_{1}$
and $\ell_{2}$ \citep{Kaiser:1992ps}.

For a given source galaxy, one considers 
the relation between the observed ellipticity 
$\epsilon_{{\rm obs},\alpha}$ and the expected shear 
$\tilde{\gamma}_{\alpha}$,
\beqa
\tilde{\gamma}_{\alpha} 
&=& \frac{\epsilon_{{\rm obs},\alpha}}{2\cal{R}} ,\\
\tilde{\gamma}_{\alpha}
&=& (1+m_{\rm b}) \gamma_{{\rm true}, \alpha} + c_{\alpha}, \label{eq:mbias}
\eeqa
where $\cal{R}$ is the conversion factor to represent 
the response of the distortion of the galaxy image to 
a small shear \citep{Bernstein:2001nz},
$\gamma_{{\rm true}, \alpha}$ is the true value of cosmic shear,
and $m_{\rm b}$ and $c_{\alpha}$ are
the multiplicative and additive biases that represent possible 
systematic uncertainty in galaxy shape measurements. In practice,
before employing the conversion in Eq.~(\ref{eq:gamma2kappa}),
one must first construct a smoothed shear field 
on grids \citep{Seitz:1994gz},
\beqa
\gamma_{{\rm grid}, \alpha}(\bd{\theta}) 
&=& 
{\scriptsize
\frac{\sum_{i\in \bd{\theta}} w_{i} 
\left(\epsilon_{i, {\rm obs}, \alpha}/2{\cal R}-c_{i, \alpha}\right)}{(1+\langle m_{\rm b} \rangle) \sum_{i\in \bd{\theta}}\, w_{i}}}, \label{eq:grid_shear} \\
\langle m_{\rm b} \rangle &=& \frac{\sum_{i\in {\rm all}} w_{i} m_{{\rm b}, i}}{\sum_{i\in {\rm all}} w_{i}}, \label{eq:mbias_for_single_pop} \\
\gamma_{{\rm sm}, \alpha}(\bd{\theta}) &=& \int {\rm d}^{2} \phi \, \gamma_{{\rm grid}, \alpha}(\bd{\phi}) W(\bd{\phi}-\bd{\theta}) 
\label{eq:smoothed_shear}
\eeqa
where $\bd{\theta}_{i}$ is the position of the $i$-th 
source galaxy, $w_{i}$ represents the inverse variance weight, 
and $W(\bd{\theta})$ is 
a smoothing filer. In the above, $\sum_{i \in {\scriptsize \bd{\theta}}}$ represents
the summation over the galaxies in the pixel at 
the angular coordinate $\bd{\theta}$, while $\sum_{i \in {\rm all}}$ is the sum over all the galaxies in our survey window.
In this paper, we assume the functional form for $W$ as
\beqa
W(\bd{\theta}) = \frac{1}{\pi \theta^2}\left[ 1-\left(1+\frac{\theta^2}{\theta^2_{s}}\right)\exp\left(-\frac{\theta^2}{\theta^2_s}\right)\right],
\label{eq:filter_for_shear}
\eeqa
for $\theta\le10\, \theta_{s}$ and $W(\bd{\theta})=0$ otherwise.
We set $\theta_{s}=6\, {\rm arcmins}$ throughout this paper\footnote{
The smoothing scale $\theta_{s}$ is commonly adopted to search for massive galaxy clusters in a smoothed lensing map \citep{Hamana:2003ts}.
Using numerical simulations, \citet{2015MNRAS.453.3043S} has found a one-to-one correspondence between the peaks on a smoothed map by the filter in Eq.~(\ref{eq:filter_for_shear}) and massive galaxy clusters at $z=0.1-0.3$ when imposing the peak height to be larger than $\sim5\sigma$.}.
Using Eqs.~(\ref{eq:gamma2kappa}) and (\ref{eq:smoothed_shear}),
one can derive the smoothed convergence field from the observed imaging 
data through Fast Fourier Transform (FFT).

The observed ellipticity can be expressed as a sum of two term in practice:
\beqa
\epsilon_{\rm obs} = {2 \cal R} \, \gamma + \epsilon_{\rm N},
\eeqa
where $\gamma$ is the lensing shear of interest and $\epsilon_{\rm N}$ represents  noise that originates from the intrinsic galaxy shape and from 
observational conditions, referred to as shape noise.
Accordingly, we have two components in the observed lensing map as
\beqa
\kappa_{\rm obs} = \kappa_{\rm WL} + \kappa_{\rm N}.
\eeqa
The shape noise is much larger than the lensing shear term for individual objects 
in typical galaxy imaging surveys. Hence, the observed map $\kappa_{\rm obs}$ is 
 contaminated
by the shape noise on a pixel-by-pixel basis, which makes it challenging to
extract the cosmological information contained in the map.
Our objective in this paper is to estimate the noiseless field $\kappa_{\rm WL}$ from 
the observed (noisy) map $\kappa_{\rm obs}$. For this purpose, 
we use conditional generative adversarial networks (GANs).




\section{Data}
\label{sec:data}

\subsection{Subaru Hyper Suprime-Cam Survey} 

Hyper Suprime-Cam (HSC) is a wide-field imaging camera 
installed at the prime focus of the 8.2-meter Subaru telescope
\citep{Miyazaki:2015haa, Aihara:2017paw, 2018PASJ...70S...2K, 2018PASJ...70S...3F, 2018PASJ...70S...1M}. 
The Wide Layer in the HSC survey will cover 1400 ${\rm deg}^2$ 
in five broad photometric bands ($grizy$) 
in its 5-year operation, with superb image quality of sub-arcsec seeing.   
In this paper, we use a galaxy shape catalogue
that has been produced for cosmological weak lensing analysis in the first year data release (HSC S16A hereafter). Details of the galaxy shape measurements and catalogue information are found in \citet{2018PASJ...70S..25M}.

In brief, the HSC S16A galaxy shape catalogue is made from the HSC Wide-Layer data taken from March 2014 to April 2016 over 90 nights. 
We use the same set of galaxies as in \citet{2018PASJ...70S..25M}
to construct a ``secure'' shape catalogue for weak lensing analysis. 
The sky areas around bright stars are masked \citep{2018PASJ...70S...7C}. 
The HSC S16A weak lensing shear catalogue covers 136.9~deg$^2$ 
that consists of the following 6 disjoint patches: XMM, GAMA09H, GAMA15H, HECTOMAP, VVDS, and WIDE12H. 
Among the 6 patches, we choose the XMM field as a main sample in this paper 
because there exist publicly available catalogues of galaxy clusters in optical \citep{Oguri:2017khw} and X-ray bands \citep{Adami:2018ysh}.
We can use the cluster catalogues to examine the reliability of our denoising process by performing object-by-object matching.

In the HSC S16A shape catalogue, the galaxy shapes  
are estimated 
using the re-Gaussianisation PSF correction method 
applied to the $i$-band coadded images \citep{Hirata:2003cv}. 
  
In the XMM region, the survey window is defined such that 1) the number of visits within {\tt HEALPix} pixels with {\tt NSIDE=1024} to be $(g,r,i,z,y)\geq(4,4,4,6,6)$ and the $i$-band limiting magnitude to be greater than 25.6, 2) the PSF modelling is 
sufficiently good to meet our requirements on PSF model size residuals and residual shear correlation functions, 3) there are no disconnected {\tt HEALPix} pixels after the cut 1) and 2), and 4) the galaxies do not lie within the bright object masks. For details of defining these masks, see \citet{2018PASJ...70S..25M}.

\begin{figure}
\begin{center}
       \includegraphics[clip, width=1.\columnwidth]
       {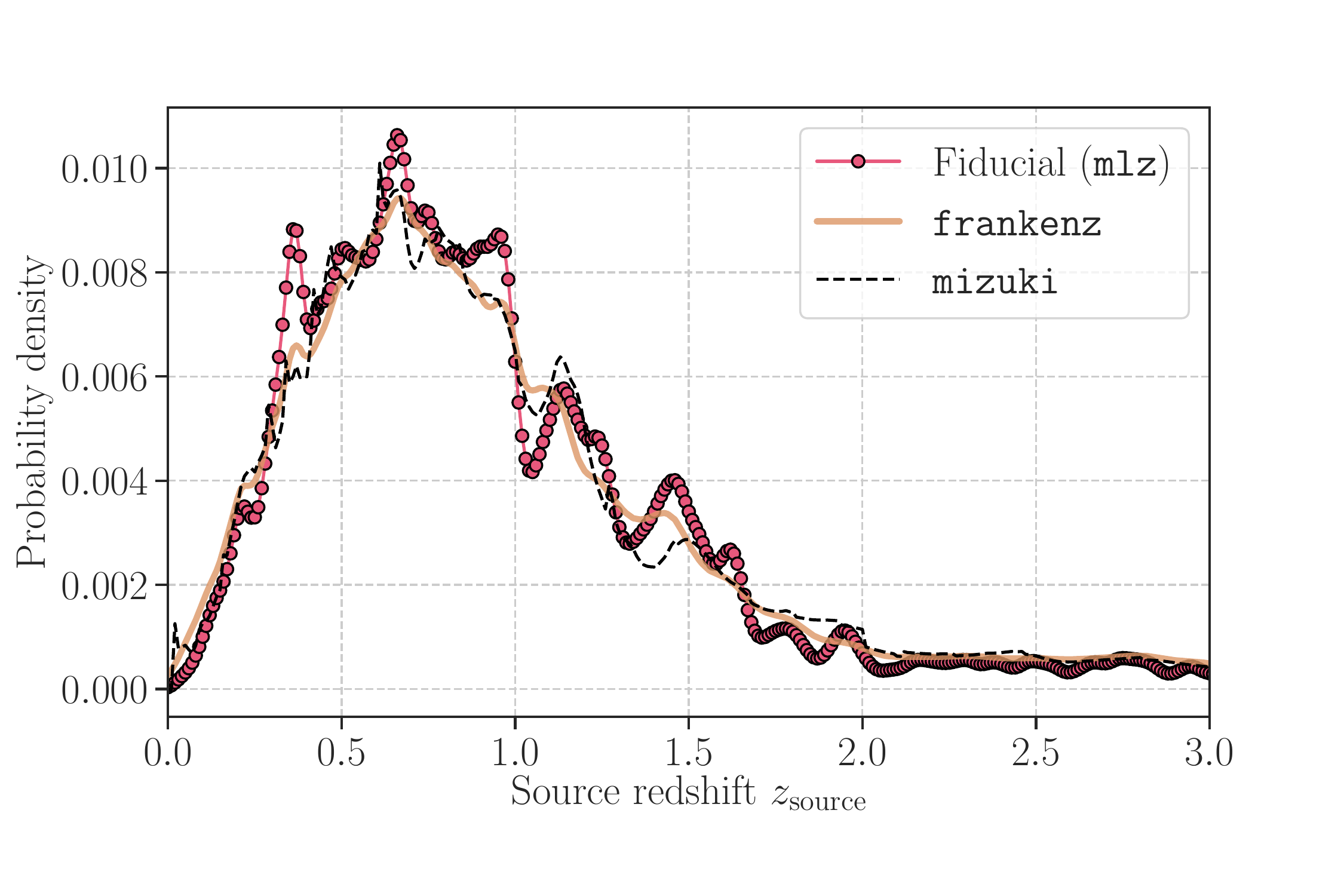}
     \caption{
     \label{fig:pz}
     The stacked photometric redshift distribution for the galaxies in the XMM field.
     The line with points shows the estimate by our baseline method, while the yellow solid
     and black dashed lines stand for the results by {\tt frankenz} and {\tt mizuki}, respectively.
  } 
    \end{center}
\end{figure}

The redshift distribution of the source galaxies 
is estimated from the HSC five broadband photometry. 
\citet{2018PASJ...70S...9T} measure photometric redshifts (photo-$z$'s)
of the galaxies in the HSC survey by using several different methods.
Among them, we choose the photo-$z$ with a machine-learning code based on self-organising map 
({\tt mlz}) as a baseline.
To study the impact of photo-$z$ estimation with different methods, 
we consider two additional photo-$z$'s estimated 
from a classical template-fitting code ({\tt mizuki})
and a hybrid code combining machine learning with template fitting ({\tt frankenz}). 
For our analysis, we select the source galaxies by their {\it best} estimates (see \citet{2018PASJ...70S...9T}) of the photo-$z$'s ($z^{\rm best}$) 
in the redshift range from 0.3 to 1.5 as done in the main cosmological analyses for the HSC S16A data \citep{Hikage:2018qbn}.
For a given method of the photo-$z$ estimation, 
individual HSC galaxies are assigned
a posterior probability distribution function (PDF) of redshift.
Figure~\ref{fig:pz} shows the stacked PDFs for the source galaxies in the XMM.
The mean source redshifts are found to be 0.96, 1.01, and 1.01 for the estimates 
by {\tt mlz}, {\tt mizuki}, and {\tt frankenz}, respectively.

We then reconstruct the smoothed convergence field from 
the HSC S16A data as described in Section~\ref{subsec:reconst_kappa}.
Adopting a flat-sky approximation, 
we first create a pixelised shear map for 
the XMM on regular grids with a grid size of 1.5 arcmins. 
We then apply FFT and perform 
convolution in Fourier space to obtain the smoothed convergence field.
Note that we limit the maximum number of grids on a side to be 256 
in our analysis.
Currently, it is still computationally expensive to train GANs with 
large-size images with decent computer resources (see \citet{2018arXiv180911096B} 
for a recent attempt).
Since our aim here is to analyse lensing convergence maps 
with an arcmin resolution,
the pixel size is set to be $\sim1\, {\rm arcmin}$. 
We will analyse observational data for a larger region
in our future work.
Our survey window covers the range of 
$[30.9, 37.3]\, {\rm deg}$ and 
$[-7.29, -0.89]\, {\rm deg}$ in right ascension (RA)
and declination (dec), respectively.
There are 1345810 source galaxies available with the photo-$z$ estimate by {\tt mlz}.
\rev{On the other hand, we found 1345541 and 1342017 objects
for the selection based on {\tt mizuki} and {\tt frankenz}, respectively.
Note that the selection of source galaxies depends on how to estimate the photo-$z$.}

In actual observations, there are missing galaxy shear data 
due to bright star masks. The observed regions
have also complex geometry. 
Applying our method directly to such regions likely generates
additional noises \citep{2013ApJ...774..111S}. 
We determine the mask regions for each convergence map
by using the smoothed number density map of the input galaxies
with the same smoothing kernel as in Eq.~(\ref{eq:filter_for_shear}). 
Then we mask all the pixels with the smoothed
galaxy number density less than 0.5 times the mean number density.
After masking, the data region is found to cover 21.4 ${\rm deg}^2$.

\begin{table*}
\caption{
\label{tab:HSC_mock_summary} 
Summary of our mock catalogues for Subaru Hyper-Suprime Cam Survey first-year data.
For each of 100 cosmological models (parameter sets), we have 50 realisations of mock catalogues.
}
\begin{tabular}{|c|c|c|c|c|}
\hline
Name & 
\verb|#| of realisations &
Cosmology &
Note  &
Reference \\ \hline
Fiducial
& 2268 & WMAP9 cosmology \citep{Hinshaw:2012aka} & Photo-$z$ info by {\tt mlz} & Section~\ref{subsubsec:mock_fid} \\
\hline
Photo-$z$ run 1
& 100 & - & Photo-$z$ info by {\tt mizuki} & Section~\ref{subsubsec:mock_photoz} \\
Photo-$z$ run 2
& 100 & - & Photo-$z$ info by {\tt frankenz} & - \\
\hline
multiplicative-bias run 1
& 100 & - & Change $\langle m_{\rm b} \rangle$ by $+0.01$ & Section~\ref{subsubsec:mock_mbias} \\ 
multiplicative-bias run 2
& 100 & - & Change $\langle m_{\rm b} \rangle$ by $-0.01$ & - \\ 
\hline
Noise-varied run 1 & 100 & - & 
Change $\sigma_{\rm mea}$ by $+10\%$ & Section~\ref{subsubsec:mock_sigma_mea} \\ 
Noise-varied run 2 & 100 & - & 
Change $\sigma_{\rm mea}$ by $-10\%$ & - \\ 
\hline
Cosmology-varied run
& 50$\times$100 & 100 different models (Figure~\ref{fig:om_S8}) & Photo-$z$ info by {\tt mlz} & Section~\ref{subsubsec:mock_cosmology} \\
\hline
\end{tabular}
\end{table*}

\subsection{Mock HSC observations}

We use a large set of simulation data for training our conditional GANs.
Table~\ref{tab:HSC_mock_summary} summarises our mock simulations.

\subsubsection{Fiducial simulations}
\label{subsubsec:mock_fid}

We first describe the mock shape catalogues for HSC S16A.
The mock catalogues are generated from 108 full-sky lensing simulations presented in \citet{2017ApJ...850...24T}\footnote{The full-sky light-cone simulation data are freely available for download at \url{http://cosmo.phys.hirosaki-u.ac.jp/takahasi/allsky_raytracing/}.}.
In \citet{2017ApJ...850...24T},
the authors perform a suite of cosmological $N$-body simulations with $2048^3$ particles and generate lensing convergence maps and halo catalogues. 
The $N$-body simulations assume the standard $\Lambda$CDM cosmology 
consistent with the 9-year WMAP cosmology (WMAP9)
\citep{Hinshaw:2012aka} with
the CDM density parameter 
$\Omega_{\rm cdm}=0.233$, 
the baryon density 
$\Omega_{\rm b0}=0.046$, 
the matter density 
$\Omega_{\rm m0}=\Omega_{\rm cdm}+\Omega_{\rm b0}
= 0.279$, 
the cosmological constant 
$\Omega_{\Lambda}=0.721$, 
the Hubble parameter
$h= 0.7$, 
the amplitude of density fluctuations
$\sigma_8= 0.82$,
and the spectral index
$n_s= 0.97$.
The gravitational lensing effect is simulated with the multiple lens-plane algorithm on a curved sky \citep{2013MNRAS.435..115B,2015MNRAS.453.3043S}.  Light-ray deflection is directly followed by using the projected matter density field 
produced by the outputs 
from the $N$-body simulations. Each lensing simulation data consists of 38 different source planes at redshift less than 5.3. Realistic source redshift distributions
are implemented following the curves in Figure~\ref{fig:pz}.

To generate mock shape catalogues, we employ essentially the same method 
as developed in \citet{Shirasaki:2013zpa, Shirasaki:2016fuf}.
We use the full-sky simulations combined with
the observed photometric redshifts and angular positions of real galaxies. 
Provided the real catalogue of source galaxies, 
where each galaxy contains information on the position (RA and Dec), shape, redshift, and the lensing weight, 
we perform the following four-steps:

\begin{description}
 \item[(i)] Set the RA and Dec of the survey window in the full-sky realisation.
 \item[(ii)] Populate source galaxies on the light-cone using original angular positions and redshifts of the observed galaxies. 
 \item[(iii)] Rotate the shape of each source galaxy at random to erase the real lensing signal.
 \item[(iv)] Add the lensing shear on each source galaxy using the lensing simulations
\end{description}

\rev{In the step (ii), we draw the source redshift at random by following the posterior distribution of photo-$z$ estimates on an object-by-object basis. Hence, our mock catalogue contains galaxies at $z<0.3$ or $z>1.5$.} 
Note that our method maintains the observed properties of 
the source galaxies on the sky.
We increase the number of realisations of the mock catalogues 
by extracting multiple separate regions from a single full-sky simulation.
Finally we obtain 2268 mock catalogues in total\footnote{The mock shape catalogues are publicly available at \url{http://gfarm.ipmu.jp/~surhud/}.}.

\subsubsection{Photometric redshift uncertainties}
\label{subsubsec:mock_photoz}

In the fiducial mock catalogues, we utilise the photo-$z$ information estimated by {\tt mlz}.
To examine possible systematic effects owing to photo-$z$ uncertainties, we generate additional mock realisations 
adopting the two other redshift estimates by {\tt mizuki} or {\tt frankenz}.
We produce 100 mock realisations of the HSC S16A catalogues for each model,
and use them to evaluate the impact of photo-$z$
uncertainty in our denoising process.

\subsubsection{Image calibration uncertainties}
\label{subsubsec:mock_mbias}

We use a single value of multiplicative bias $\langle m_{\rm b} \rangle$ (defined in Eq.~[\ref{eq:mbias_for_single_pop}])
when generating our fiducial mock catalogues. 
Estimating $\langle m_{\rm b} \rangle$ is based on image simulations, and thus 
there remains a $1\%$-level uncertainty \citep{Mandelbaum:2017ctf}. 
To account for possible systematic effects by the mis-estimation 
of the multiplicative bias,
we make additional mock realisations by changing 
$\langle m_{\rm b} \rangle \rightarrow \langle m_{\rm b} \rangle + \Delta m_{\rm b}$
in the production process. We assume two values of $\Delta m_{\rm b} = \pm0.01$.
For each value of $\Delta m_{\rm b}$, we produce 100 mock realisations of the HSC S16A.

\subsubsection{Noise model uncertainties}\label{subsubsec:mock_sigma_mea}
Imperfect knowledge of the noise distribution in the data 
can be another source of systematic uncertainties 
in denoising process.
To assess possible model bias in the noise, we generate two different mock 
catalogues by varying the amplitude of the standard deviation (error) of the shape measurement, $\sigma_{\rm mea}$.
In the HSC S16A data, the value of $\sigma_{\rm mea}$ has been calibrated with a set of image simulations, and the estimate for individual objects may be subject to a $10\%$-level uncertainty \citep{Mandelbaum:2017ctf}. 
To test the impact of this uncertainty, we vary the amplitude of $\sigma_{\rm mea}$ 
by a factor of $(1+\Delta \sigma_{\rm mea})$ 
on an object-by-object basis when generating mock catalogues.
\rev{We keep the lensing weight fixed even when varying $\sigma_{\rm mea}$ in the lensing analysis, because we suppose that we are unaware of the mis-estimate of $\sigma_{\rm mea}$.}
We assume two values, $\Delta \sigma_{\rm mea} = 0.1$ and $\Delta \sigma_{\rm mea} = -0.1$.
For each value of $\Delta \sigma_{\rm mea}$, we produce 100 mock realisations of the HSC S16A.

\subsubsection{Varying cosmological models}
\label{subsubsec:mock_cosmology}

\begin{figure}
\begin{center}
       \includegraphics[clip, width=1.\columnwidth]
       {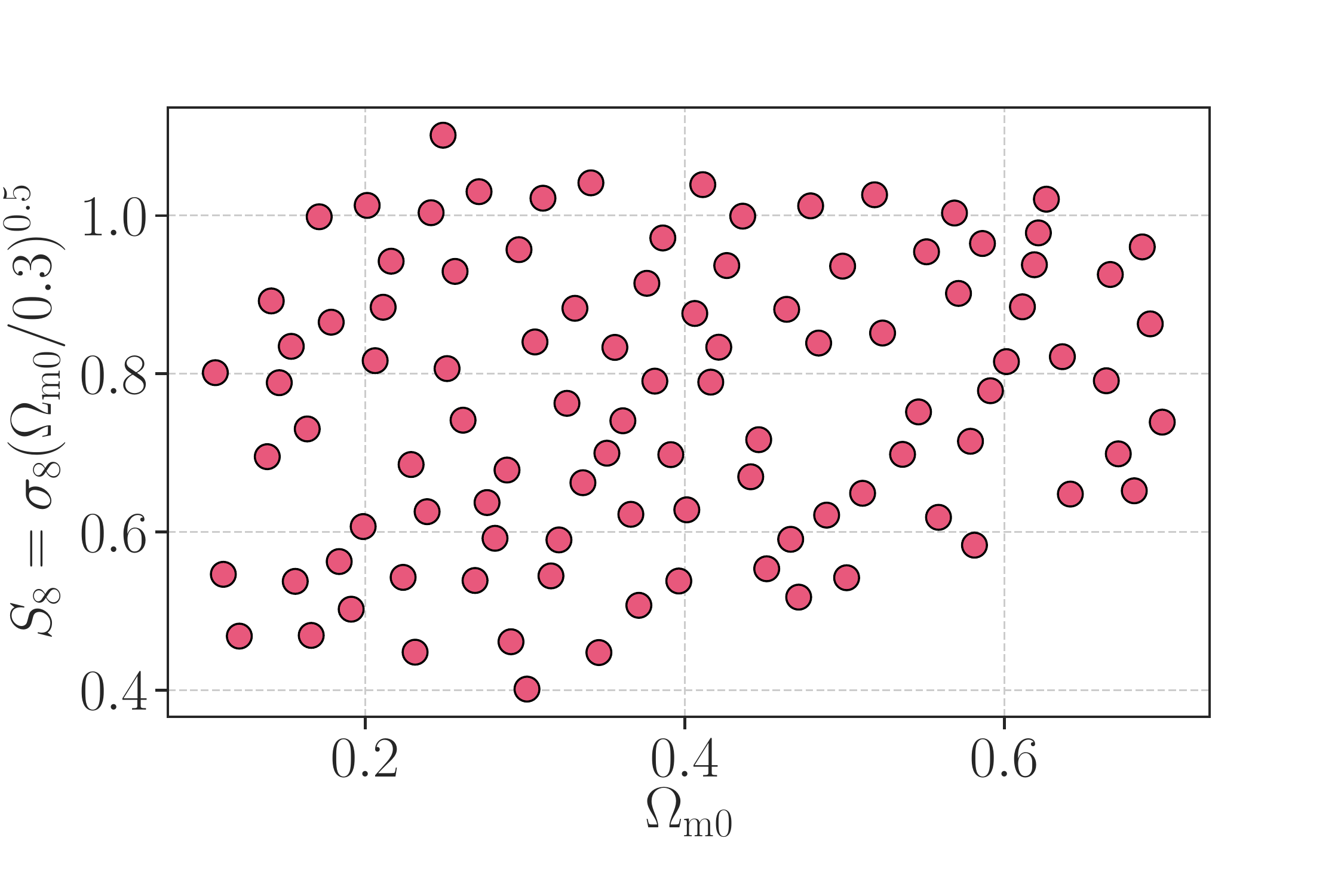}
     \caption{
     \label{fig:om_S8}
     The 100 different cosmological models to study the cosmological dependence on weak lensing maps.
     At each point, we generate 50 mock realisations of the HSC S16A data.
  } 
    \end{center}
\end{figure}

To study the cosmological dependence on weak lensing maps, we also generate mock catalogues
of the HSC S16A data by varying cosmological models.
We design the cosmological models for simulations 
so as to cover a much wider area in the two-parameter space $(\Omega_{\rm m0}, \sigma_8)$
than the constraints by the current galaxy imaging surveys \citep{Hildebrandt:2016iqg, Troxel:2017xyo, Troxel:2018qll, Hikage:2018qbn}.
We choose a sample of cosmological models in the $\Omega_{\rm m0}-\sigma_8$ plane
by using a public R package to generate
the maximum-distance sliced Latin Hypercube Designs (LHDs) \citep{SLHDs}.
\rev{We first generate 120 designs in a two dimensional rectangle specified by $0.1\le \Omega_{\rm m0} \le 0.7$ and $0.4\le \sigma_{8} (\Omega_{\rm m0}/0.3)^{0.6}\le 1.1$ using the codes. We then restrict the designs to those with $0.4\le\sigma_{8}\le1.4$. This leaves 100 designs. }
Figure~\ref{fig:om_S8} shows 
the resultant 100 cosmological models adopted in our simulations.
Note that we set $\Omega_{\Lambda} = 1-\Omega_{\rm m0}$ assuming a spatially flat universe.
For other parameters, we adopt $\Omega_{\rm b0} h^2 = 0.02225$, $h=0.6727$ and $n_s=0.9645$.
These parameters are consistent with the results from Planck 2015 \citep{Ade:2015xua}.

\begin{figure}
\begin{center}
       \includegraphics[clip, width=0.8\columnwidth]
       {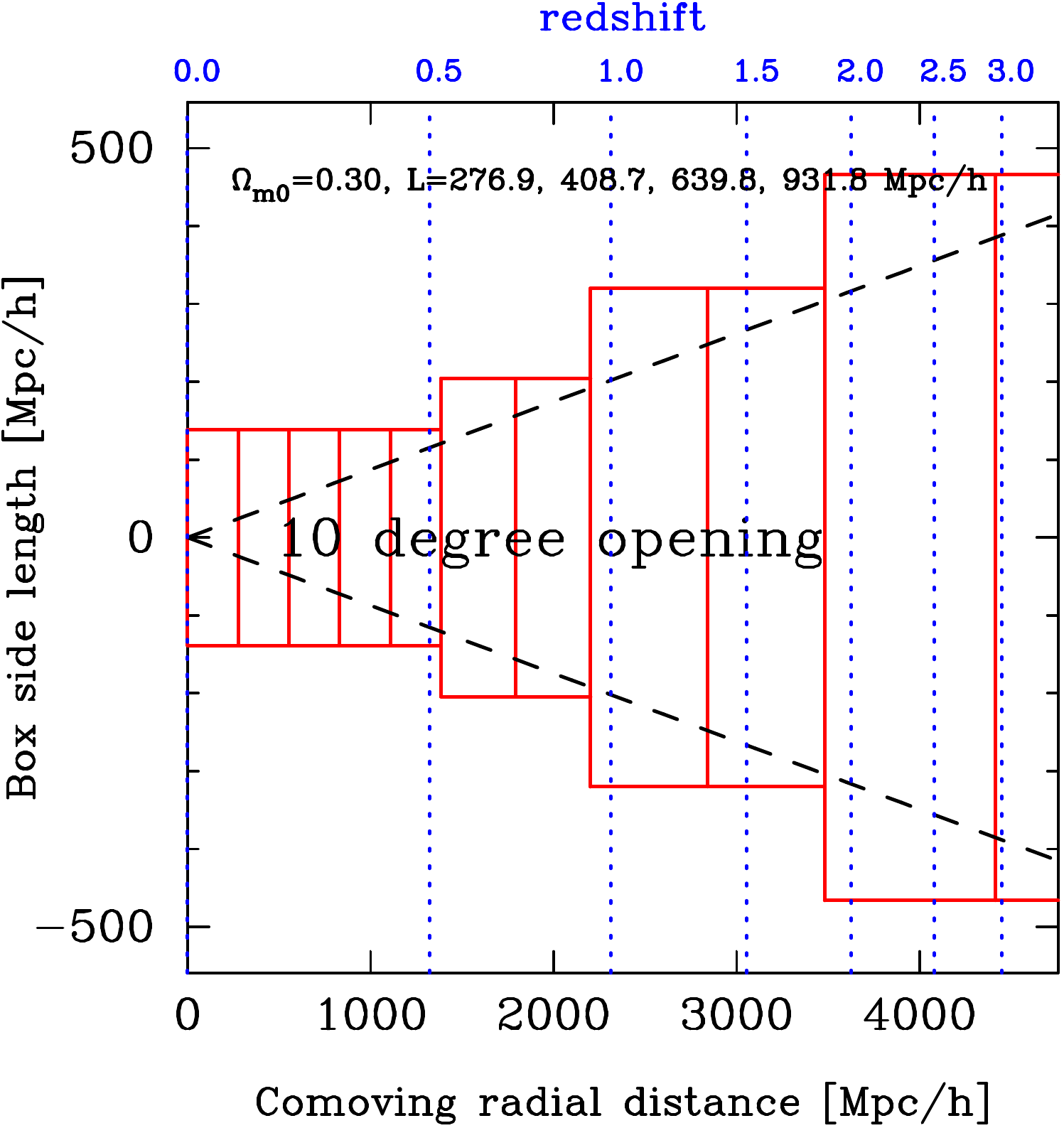}
     \caption{
     \label{fig:4box}
     The configuration of $N$-body boxes in our ray-tracing simulation for the cosmology 
     with $\Omega_{\rm m0}=0.3$.
  } 
    \end{center}
\end{figure}

For each cosmological model, we perform ray-tracing simulations under a flat-sky approximation.
We adopt the multiple lens-plane algorithm \citep{Jain:1999ir, Hamana:2001vz}
to simulate the gravitational lensing effects on a light cone of angular size $10^{\circ}\times10^{\circ}$.
We place a set of $N$-body simulations with different volumes to cover a wide redshift range 
as well as have higher mass and spatial resolutions at lower redshifts (e.g. see, \citet{Sato:2009ct}).
We consider four different box sizes on a side and each box size is varied as a function of the cosmological model.
The box size $L_{\rm box}$ of the $N$-body simulations for our ray-tracing simulations is 
set by the following criteria:
\beqa
L_{{\rm box}, 1} &=& \chi(z=0.5) \times (\theta_{\rm sim}+\Delta \theta) , \\
L_{{\rm box}, 2} &=& \chi(z=0.8) \times (\theta_{\rm sim}+\Delta \theta), \\
L_{{\rm box}, 3} &=& \chi(z=1.5) \times(\theta_{\rm sim}+\Delta \theta), \\
L_{{\rm box}, 4} &=& \chi(z=3.0) \times (\theta_{\rm sim}+\Delta \theta),
\eeqa
where $L_{{\rm box},i}$ is the box size for 
the $i$-th smallest-volume simulation and $\theta_{\rm sim} = 10\, {\rm deg}$.
We introduce the buffer in opening angle to compute $L_{\rm box}$ and set $\Delta \theta = 2\, {\rm deg}$.
We then place the $N$-body simulations with 
the box size of $L_{{\rm box}, 1}, L_{{\rm box}, 2}, L_{{\rm box}, 3}$ and $L_{{\rm box}, 4}$
to cover the light cone in the redshift range of 
$0<z<0.5$, $0.5<z<0.8$, $0.8<z<1.5$, and $1.5<z<3.0$, respectively.
Figure~\ref{fig:4box} shows an example of the configuration of $N$-body boxes in our ray-tracing simulation
in the case of $\Omega_{\rm m0}=0.3$.

For a given single $N$-body simulation volume, we produce two sets of the projected density fields with a projection depth 
of $L_{\rm box}/2$ on $9600^2$ grids by using the triangular-shaped cloud assignment scheme \citep{1992nrfa.book.....P}. By solving the discretised lens equation numerically, we obtain the lensing convergence $\kappa$ and shear $\gamma$ 
on $4096^2$ grids with a grid size of 0.15 arcmin.
A single realisation of our ray-tracing data consists of 22 source planes 
in the range of $z\simlt3$.
We perform 50 ray-tracing realisations of the underlying density field 
by randomly shifting the simulation boxes assuming periodic boundary conditions.
We finally produce the mock 
catalogue of the HSC S16A 
as described in Sec~\ref{subsubsec:mock_fid}.

When running cosmological $N$-body simulations, we use the parallel Tree-Particle Mesh code {\tt GADGET2} \citep{Springel:2005mi}. We generate the initial conditions using a parallel code developed by \citet{Nishimichi:2008ry, 2011A&A...527A..87V}, which employ the second-order Lagrangian perturbation theory \citep{Crocce:2006ve}.
The number of $N$-body particles is set to $512^3$. We set the initial redshift by 
$1+z_{\rm init} = 36 \, (512/L_{\rm box})$, where we compute the linear matter transfer function using {\tt CAMB} \citep{Lewis:1999bs}. 
Note that our choice of the initial redshift is motivated  by the detailed 
study of \citet{2019ApJ...884...29N}.


\section{Denoising by deep-learning networks}
\label{sec:analysis}


\subsection{Conditional generative adversarial networks}

To perform mapping from a noisy lensing field $\kappa_{\rm obs}$ to 
a noiseless counterpart $\kappa_{\rm WL}$,
we use a model of conditional generative adversarial networks developed in \citet{2016arXiv161107004I}.
The networks have two main components, a generator and a discriminator. 
We train the networks so that 
the generator applies some transformation to the input noisy field $\kappa_{\rm obs}$ 
to output a {\it noise} field $\kappa_{\rm N}$\footnote{\rev{One may think that it would be more appropriate to directly generate a noiseless lensing field in the network. This possibility has been examined in our previous work and it does not work in an HSC-like imaging survey \citep{Shirasaki:2018thk}. This is mainly because the signal-to-noise ratio on a pixel-by-pixel basis
is small in general.}}.
The discriminator compares the input image to an unknown image 
(either a target image from the data set 
or an output image from the generator) 
and tries to judge if it is produced 
by the generator.
To be specific, the input image for the discriminator is set 
to the noisy field $\kappa_{\rm obs}$, 
while the target image is either of the noise counterpart of $\kappa_{\rm obs}$ or an output from the generator.

The structure of the generator and the discriminator in our networks is essentially
the same as in \citet{Shirasaki:2018thk}, except for minor parameter tuning.
The generator uses a U-Net structure \citep{2015arXiv150504597R} 
with an eight set of convolution and deconvolution layers.
Each convolution layer consists of convolution with a kernel size of $5\times5$, 
the batch normalisation, and the application of the activation function of leaky ReLU with a leak slope of 0.2.
The deconvolution layer does the inverse operation of the convolution layer.
The generator also has additional skip connections between mirror layers to propagate
the small-scale information that would be lost as the size of the images decreases 
through the convolution process.
The discriminator produces a single value from a given input image 
for the decision whether the input is real or a fake.
The final output of the discriminator is made after 
the image reduction through 4 convolution layers 
and after averaging all the
responses from the convolution layers.
In the convolution layers in the discriminator, 
we remove the batch normalization to balance
the losses of the generator and the discriminator in a stable way.
The resulting number of parameters in our networks is close to 400000.

\subsection{Training the networks}

The objective of our networks is
to solve an optimization problem with a cost function
expressed as a combination of loss functions as
\beq
{\rm min}_{G}\, {\rm max}_{D}\,
\Big\{{\cal L}_{\rm cGAN}(G, D) + \lambda {\cal L}_{\rm L1}\Big\},
\label{eq:obs_pix2pix}
\eeq
where $G$ indicates the generator and $D$ is the discriminator.
We here introduce two loss functions as 
\beqa
{\cal L}_{\rm cGAN}(G, D) &=& \mathbb{E}_{x,y}\log D(x, y) \nonumber \\
&&
\,\,\,
+\mathbb{E}_{x,z}\log\left\{1-D(x, G(x,z))\right\}, \\
{\cal L}_{\rm L1}(G) &=& \mathbb{E}_{x,y,z}\, \sum_{\rm map} \left|y-G(x, z)\right|, \label{eq:L1norm}
\eeqa
where $x$ is the input noisy field, 
$y$ is the true noise field,
and $z$ is a random noise vector 
at the bottom layer of the generator.
The function $D(X_1,X_2)$ returns the score in the range of zero to unity to evaluate if the noise counterpart of $X_1$ 
and a noise field $X_2$ 
are
identical or not.
In Eq.~(\ref{eq:L1norm}), the summation runs over all the pixels in a map
but with the masked region excluded.
In the training, we alternate between one gradient descent step on $D$, then one step on $G$. As suggested in \citet{2014arXiv1406.2661G}, we train to maximise
the term of $\log D(x,G(x,z))$. Also, 
we divide the objective by 2 while optimising $D$, which slows down the learning rate of $D$ relative to $G$. 

When training the networks, we use the minibatch Stochastic Gradient Descent (SGD) method and apply the Adam solver \citep{2014arXiv1412.6980K}, with learning rate $0.0002$, momentum parameters $\beta_1=0.5$ and $\beta_2=0.9999$. We also set $\lambda=75$ in Eq.~(\ref{eq:obs_pix2pix}). 
The parameter $\lambda$ controls the strength of
the regularisation given by the L1 norm.
All the networks in this paper are trained with a batch size of 1. 
We initialise the model parameters in the networks from a Gaussian distribution with a mean $0$ and a standard deviation of $0.02$.
We train our networks using the TensorFlow implementation\footnote{We use the modified version of \url{https://github.com/yenchenlin/pix2pix-tensorflow}} 
on a single NVIDIA Quadro P5000 GPU.
While processing, we randomly select training and validation data from the input data sets.
Each network is validated every time it learns 100 image pairs. 

\begin{figure*}
\begin{center}
       \includegraphics[clip, width=2.2\columnwidth, viewport=100 50 1440 576]
       {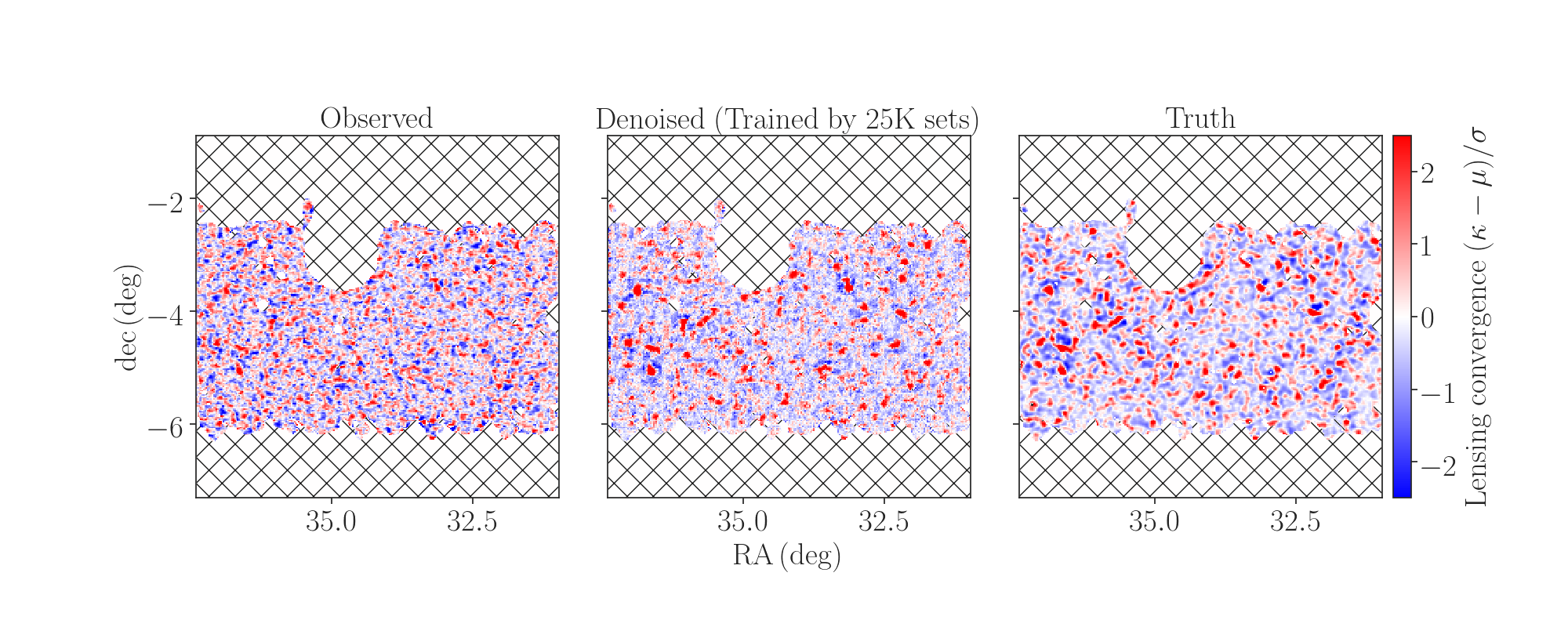}
     \caption{
     An example of image-to-image translation by our networks.
     The left panel shows an input noisy lensing map, while the right stands for the true (noiseless)
     counterpart. The medium represents the reconstructed map by our conditional GANs. 
     For the reconstructed map,
     we first obtain the underlying noise field from 10 bootstrap realisations of the generators in our GANs and then derive the convergence map by the residual between the input noisy map and the predicted noise.
     In this figure, the hatched region shows the masked area due to the presence of bright stars
     and inhomogeneous angular distributions of galaxies in our survey window.
     In the legend, $\mu$ and $\sigma$ denote the spatial average and the root-mean-square of lensing fields, respectively.
     \label{fig:train_image_fid}
  } 
    \end{center}
\end{figure*}

To prepare the training data set, we use 400 realisations of our 
mock HSC S16A catalogues (Section~\ref{subsubsec:mock_fid}).
Using the information of noiseless lensing maps $\kappa$ and $\gamma$ in our survey window, 
we generate 60000 noisy maps by injecting independent noise realizations at random.
From the 60000 image pairs of the noisy field $\kappa_{\rm obs}$ and the underlying noise $\kappa_{\rm N}$,
we select 25000 image pairs by bootstrap sampling so that 
each bootstrap realisation can contain 
167 realisations of noiseless lensing fields.
In our previous study, we find that it is near-optimal 
to use $\sim200$ realisations of noiseless lensing fields
and set the number of training sets to $\sim30000$ for our networks \citep{Shirasaki:2018thk}.

\rev{To set the hyperparameter in our network, we examined the training with $\lambda=25, 50, 75, 100$, and $150$. For a given $\lambda$, we varied the number of image pairs in the training process from 20000 to 40000 at intervals of 5000. We then apply the network trained with different hyperparameters 
to the test sets. After some trials, we find that the training with 25000 image pairs and $\lambda=75$
can provide the best performance on noise reduction in the test sets.}


\subsection{Production of the final denoised image}
\label{subsubsec:denoise_process}

As reported in \citet{Shirasaki:2018thk}, 
a single set of our networks trained by 25000 image pairs has a large scatter in the image-to-image translation. 
To reduce this dependence on training data sets, we generate 10 bootstrap sampling of 25000 training data and obtain a total 10 networks for denoising.
Namely, we obtain 10 candidates of the underlying noise field $\kappa_{\rm N}$ for a given noisy field $\kappa_{\rm obs}$. 
To evaluate the best estimate of $\kappa_{\rm N}$, we take the median over the 10 candidates on a pixel-by-pixel basis.
Once the averaged estimate of $\kappa_{\rm N}$ is determined in this manner,
we evaluate the underlying noiseless field $\kappa_{\rm WL}$
by subtracting the best noise model from the observed one $\kappa_{\rm obs}$.

The denoising process by our networks is tested 
by 1000 noisy data from the fiducial mock catalogues.
These test data are not used in the training process.

\section{Properties of denoised maps}
\label{sec:prop_denoise_map}

In this section, we study
statistical properties of weak lensing maps denoised by our conditional GANs.
We pay special attention to non-Gaussian information in the maps.
In this paper, we consider one-point distribution function (PDF)
to extract non-Gaussian information.
Furthermore, we employ matching analyses of peaks in the maps and massive clusters in 
the N-body simulations,
demonstrating that our GANs do not erase rich cosmological information from high-density regions.
In Appendix~\ref{apdx:add_tests}, we summarise additional tests for our GANs.
The tests include a conventional two-point correlation analysis, 
a reliability check of our GANs' predictions,
and dependence of our results on hyperparameter in our GANs.

Our training strategy for conditional GANs is provided in Section~\ref{sec:analysis}.
Here, we show the validation results of the outputs from our networks by using 1000 test data sets.
These test sets are based on the fiducial mock catalogues as in Section~\ref{subsubsec:mock_fid}, 
while we do not use them in the training process.
In the following, the lensing map is normalised so as to have zero mean and unit variance.

\subsection{Visual comparison}

A quick visual comparison would allow us to
highlight 
how our GAN-based denoising works for noisy input images.
Figure~\ref{fig:train_image_fid} compares three maps for one of our test data.
In the figure, the left and right panels show an input noisy and 
the true noiseless counterpart, respectively. The middle panel shows 
the denoised map by our conditional GANs.
In each panel, red spots indicate high density regions, 
while bluer ones have lower densities.
The denoised image 
retains similar patterns in density contrast over 
a few degrees compared to the ground truth.
Note that Figure~\ref{fig:train_image_fid} concentrates on the pixel values in the range of 
$-2.5\sigma$ to $+2.5\sigma$, i.e., largely noise dominated.
Although not perfect, our GANs recover
small-scale information 
(e.g.,~positive peaks) closely to the ground truth.

\subsection{One-point probability distribution}\label{subsec:1pdf}

\begin{figure}
\begin{center}
       \includegraphics[clip, width=1.15\columnwidth]
       {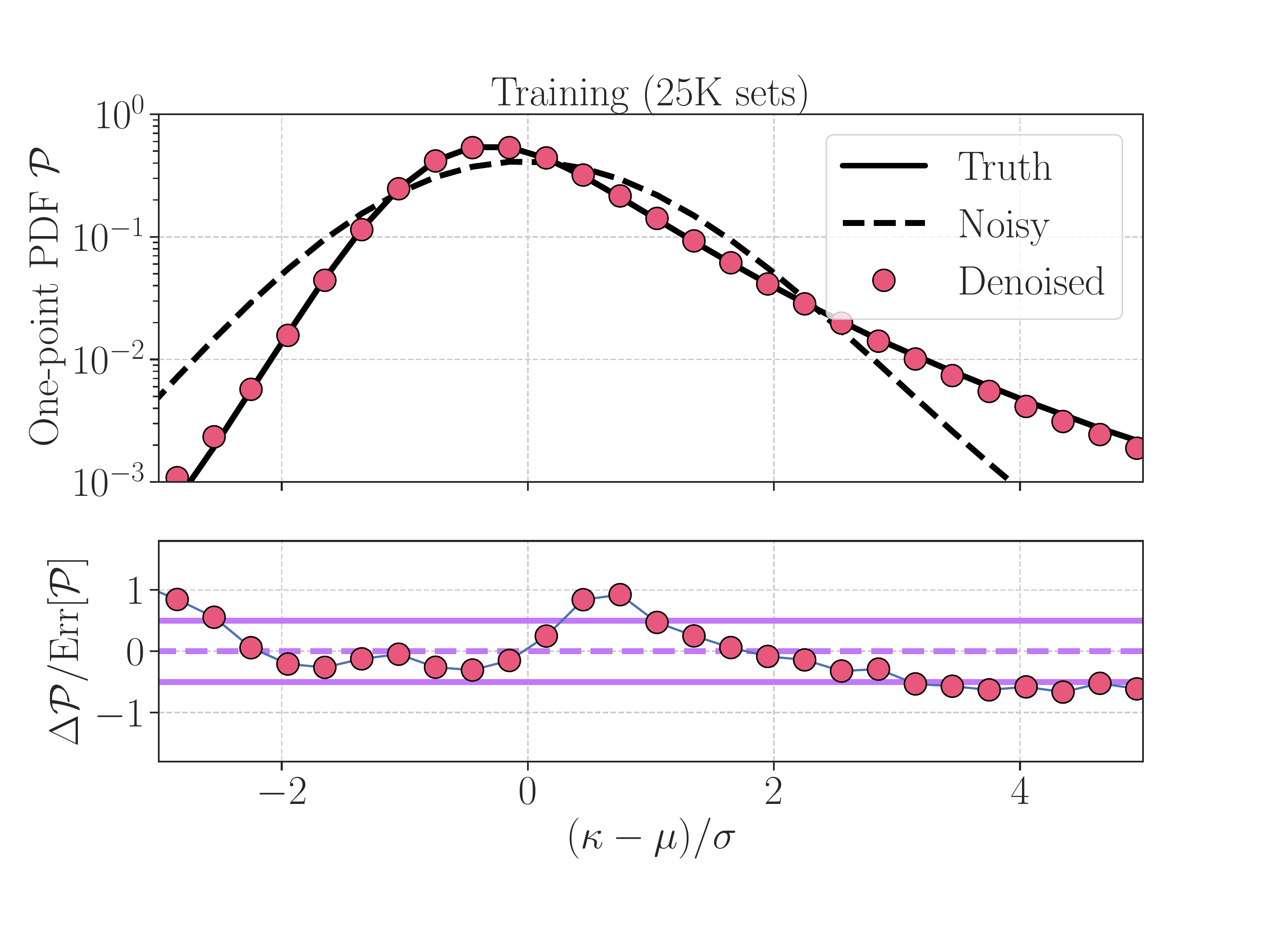}
     \caption{
     \label{fig:1pdf_comp}
     We compare the lensing PDFs for noisy, noiseless and denoised maps.
     The solid line in the top panel 
     shows the averaged PDF over the 1000 noiseless lensing maps,
     while the dashed line is for the noisy (observed) one.
     The red points in the top panel show the averaged PDF for the denoised maps by our GANs.
     In the bottom, we show the difference between the noiseless and denoised PDFs normalised by the sample variance of the noiseless PDFs.
     For reference, we highlight $\pm0.5\sigma$-level differences by the magenta lines at the bottom.
  } 
    \end{center}
\end{figure}

One-point PDF is a simple summary statistic of a weak lensing map.
Our previous study shows that the denoised image yields a similar PDF 
to the noiseless true counterpart if the lensing field is properly normalised \citep{Shirasaki:2018thk}.
Here, we repeat the previous analysis but with including various observational effects such as complex survey geometry, 
inhomogeneous galaxy distribution on a sky, 
wide redshift distribution of source galaxies, 
and variation of the weights in the analysis.
For a given lensing map, we measure the one-point PDF as a function of $(\kappa-\mu)/\sigma$ where $\mu$ and $\sigma$ are the spatial average and the root-mean-square. 
We perform linearly spaced binning in the range of 
$-15 < (\kappa-\mu)/\sigma < 15$ with width of 0.3.
Figure~\ref{fig:1pdf_comp} compares the PDFs averaged over 1000 realisations of lensing fields.
The noiseless PDF is significantly skewed compared to the observed noisy counterparts.
Our method reproduces the large skewness 
in the noiseless PDFs from the noisy input images.
The typical bias in the reconstruction ranges
from a $0.5-1\sigma$ level 
over a wide range of pixel values as shown in the bottom panel.

\subsection{Peak-halo matching}
\label{subsec:peak_halo_match}

\begin{figure*}
\begin{center}
       \includegraphics[clip, width=2.2\columnwidth]
       {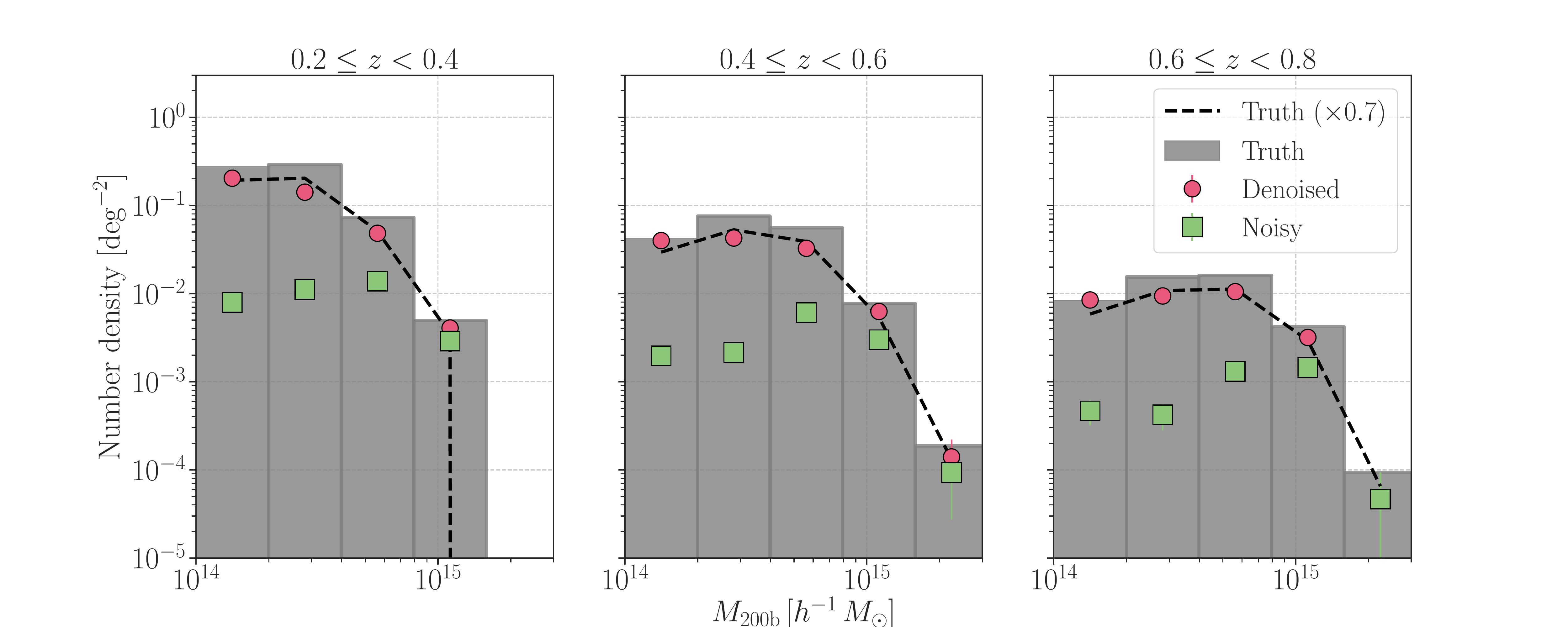}
     \caption{
     \label{fig:peak_halo_massfunc}
     The number density of the matched dark matter haloes to the peaks on the lensing peaks.
     From the left to the right, we show the number density of the dark matter haloes
     as a function of halo masses $M_{\rm 200b}$ at three different redshift ranges, 
     $0.2\le z <0.4$, $0.4\le z < 0.6$, and $0.6 \le z < 0.8$.
     The grey histogram shows the results for the true (noiseless) lensing fields,
     \rev{while the red points and green squares stand for the denoised and noisy fields, respectively.}
     The red points broadly follow the grey histogram except for the difference in amplitudes.
     For a reference, the dashed line in each panel represents the noiseless results 
     with a multiplicative factor of 0.7.
  } 
    \end{center}
\end{figure*}

\begin{figure*}
\begin{center}
       \includegraphics[clip, width=2.2\columnwidth]
       {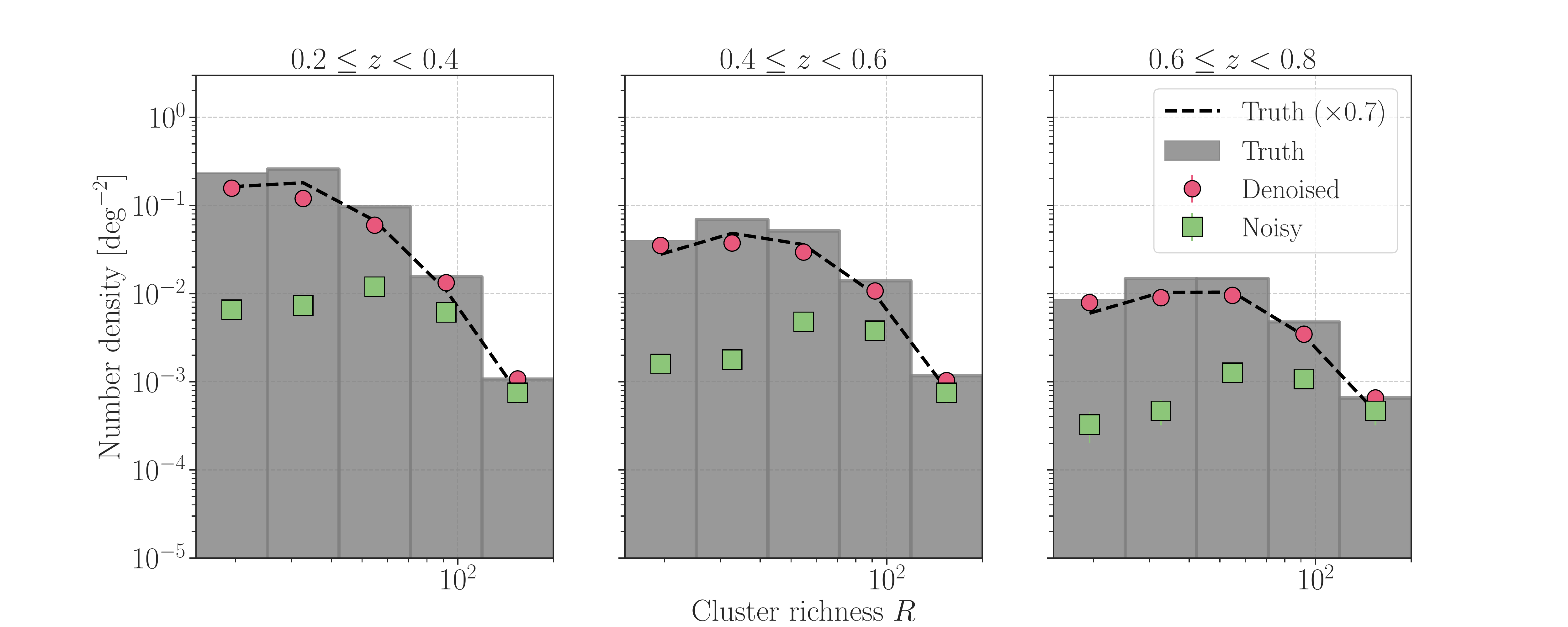}
     \caption{
     \label{fig:peak_camira_massfunc}
     Similar to Figure~\ref{fig:peak_halo_massfunc}, but we show the peak-cluster matching results for the CAMIRA-like mock catalogues.
  } 
    \end{center}
\end{figure*}

To study the small-scale structure on a denoised lensing field,
we examine the correspondence between dark matter haloes and 
the local maxima in the lensing maps.
Since our mock HSC catalogues are originally based on cosmological $N$-body simulations,
we can generate light-cone halo catalogues 
with the same sky coverage as the lensing maps. 
The light-cone catalogues are produced from the inherent full-sky halo catalogues of \citet{2017ApJ...850...24T}. The dark matter haloes in the full-sky catalogues are identified by a phase-space temporal halo finder {\tt Rockstar} \citep{2013ApJ...762..109B}.

In the following, we consider two mock cluster catalogues; one is a simple mass-limited sample, and the other takes into account a realistic mass selection effect in optically-selected galaxy clusters.
For the mass-limited sample, we use dark matter haloes with 
mass\footnote{We define the halo mass as the spherical over-density mass with respect to 200 times mean over-density.} greater than $10^{14}\, h^{-1}M_{\odot}$ at redshift less than 1.
The mass and redshift selection roughly corresponds to the real galaxy cluster catalogue based on the photometric data in HSC S16A \citep{Oguri:2017khw}.
In order to use more realistic samples, we generate mock galaxy clusters based on the multi-band identification of red sequence galaxies
\citep[the cluster finding algorithm is referred to as CAMIRA;][]{2014MNRAS.444..147O}.
We adopt the mass-to-richness relations of the CAMIRA clusters identified in HSC S16A \citep{2019PASJ...71..107M}.
\citet{2019PASJ...71..107M} assume a log-normal distribution of the cluster richness for given cluster masses and redshifts and constrain the mean and scatter relation between 
the cluster mass and richness.
Using their log-normal model, we assign the cluster richness to dark matter haloes in their redshift range of 0.1 to 1.0.

In a given lensing map, we first identify local maxima with their peak heights greater than $5\sigma$.
We then search for clusters around the peaks with a search radius of 6 arcmins. 
When we find several haloes in the search radius, 
we regard the closest cluster 
from the position of the peak as the best match.
Over 1000 realisations in our survey window, 
we find 27683 peaks. We identify $89.5\%$ of the peaks have a matched mass-limited cluster in the noiseless field.
After denoising, the number of peaks is found to be 23248 and the matching rate is $64.6\%$.
For more realistic CAMIRA-like clusters, the matching rate is found to be 
$85.1\%$ and $58.9\%$ for the noiseless and denoised fields, respectively.
\rev{Without denoising, the number of peaks reduces to 1669 over 1000 realisations.
However, the matching rate is found to be $80.2\%$ and $75.2\%$ for the mass-limited and
the CAMIRA-like clusters, respectively.
Hence, the analysis with denoising is highly complementary to the noisy counterpart.} 

Furthermore, to validate the halo-peak matching, we study the number density of the matched dark matter
haloes as a function of halo masses and redshifts.
Figures~\ref{fig:peak_halo_massfunc} and 
\ref{fig:peak_camira_massfunc}
show the number density of the matched clusters to the noiseless, the denoised, and the noisy peaks. 
As shown in the figures, the shape of the number density 
looks similar between the noiseless and the denoised peaks.
This indicates that the peak-cluster matching for the denoised fields 
is not a coincidence.
\rev{Compared to the noisy peaks, the denoised peaks play an important role to search for less massive clusters.}

\subsection{Cosmological dependence on lensing PDFs}

We next validate if our conditional GANs can reduce noises when the true cosmologocal model is different from that assumed in the training process.
When our GANs would surely learn noise properties alone, the networks should be able to denoise regardless of underlying cosmological models.
To study the cosmological dependence on the denoised lensing map, 
we use the mock catalogues as in Section~\ref{subsubsec:mock_cosmology}.
We have 50 mock realisations of noisy lensing maps for each of 100 different cosmological models.
For a given cosmology, we input a noisy map 
to our GANs, obtain the denoised map, and then compute the one-point PDF from the denoised map.
We repeat this process for 50 realisations per each cosmological model 
and estimate the average PDF.

\begin{figure}
\begin{center}
       \includegraphics[clip, width=1.1\columnwidth]
       {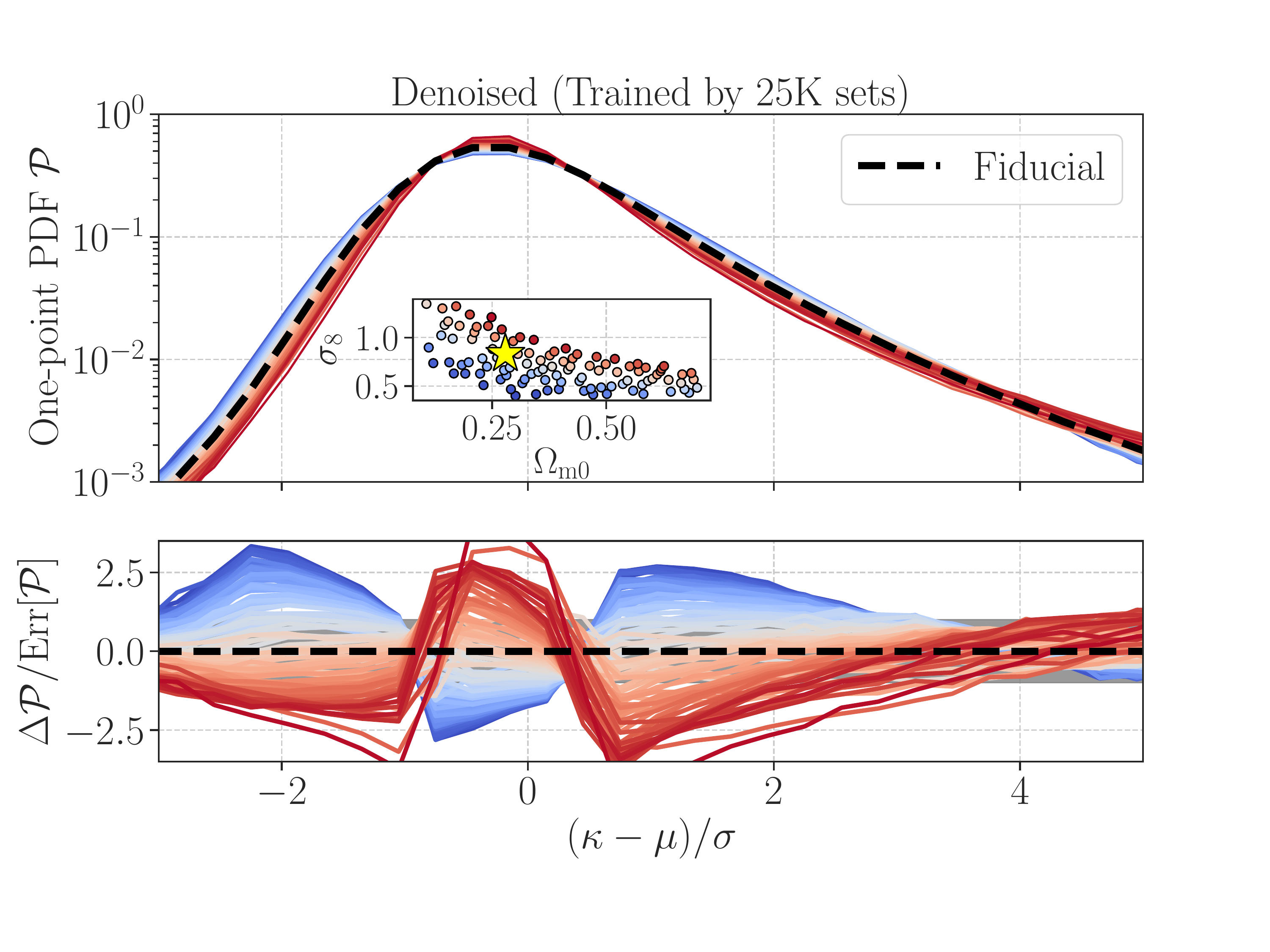}
     \caption{
     \label{fig:1pdf_denoised_cosmo}
     The cosmological dependence of the one-point PDF of the denoised maps.
     The top panel shows the PDF as a function of the pixel value $(\kappa-\mu)/\sigma$,
     where $\mu$ and $\sigma$ is the spatial average and root-mean-square for a lensing field $\kappa$, respectively. 
     In the top, the inset figure represents the 100 cosmological models considered in the present study. 
     The star symbol in the inset figure shows our fiducial model \citep[the WMAP9 cosmology;][]{Hinshaw:2012aka}.
     The dependence of cosmological models is highlighted by the colour difference.
     In the bottom, we show the difference of the PDF from our fiducial cosmological model 
     normalised by the statistical uncertainty.
  } 
    \end{center}
\end{figure}

\begin{figure}
\begin{center}
       \includegraphics[clip, width=1.1\columnwidth]
       {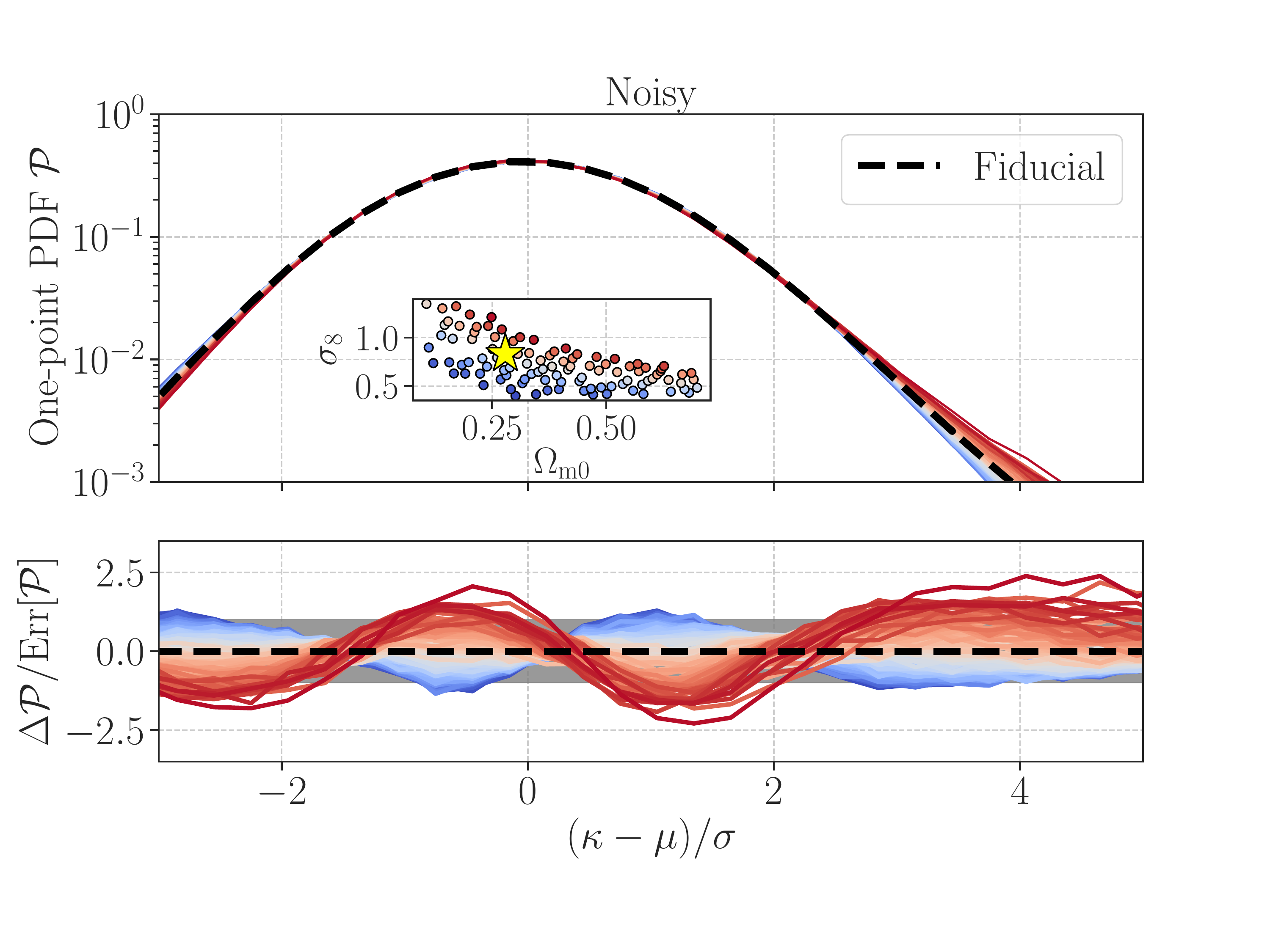}
     \caption{
     \label{fig:1pdf_noisy_cosmo}
     Similar to Figure~\ref{fig:1pdf_denoised_cosmo}, but for the lensing PDF without our denoising process.
  } 
    \end{center}
\end{figure}

\begin{figure}
\begin{center}
       \includegraphics[clip, width=1.15\columnwidth]
       {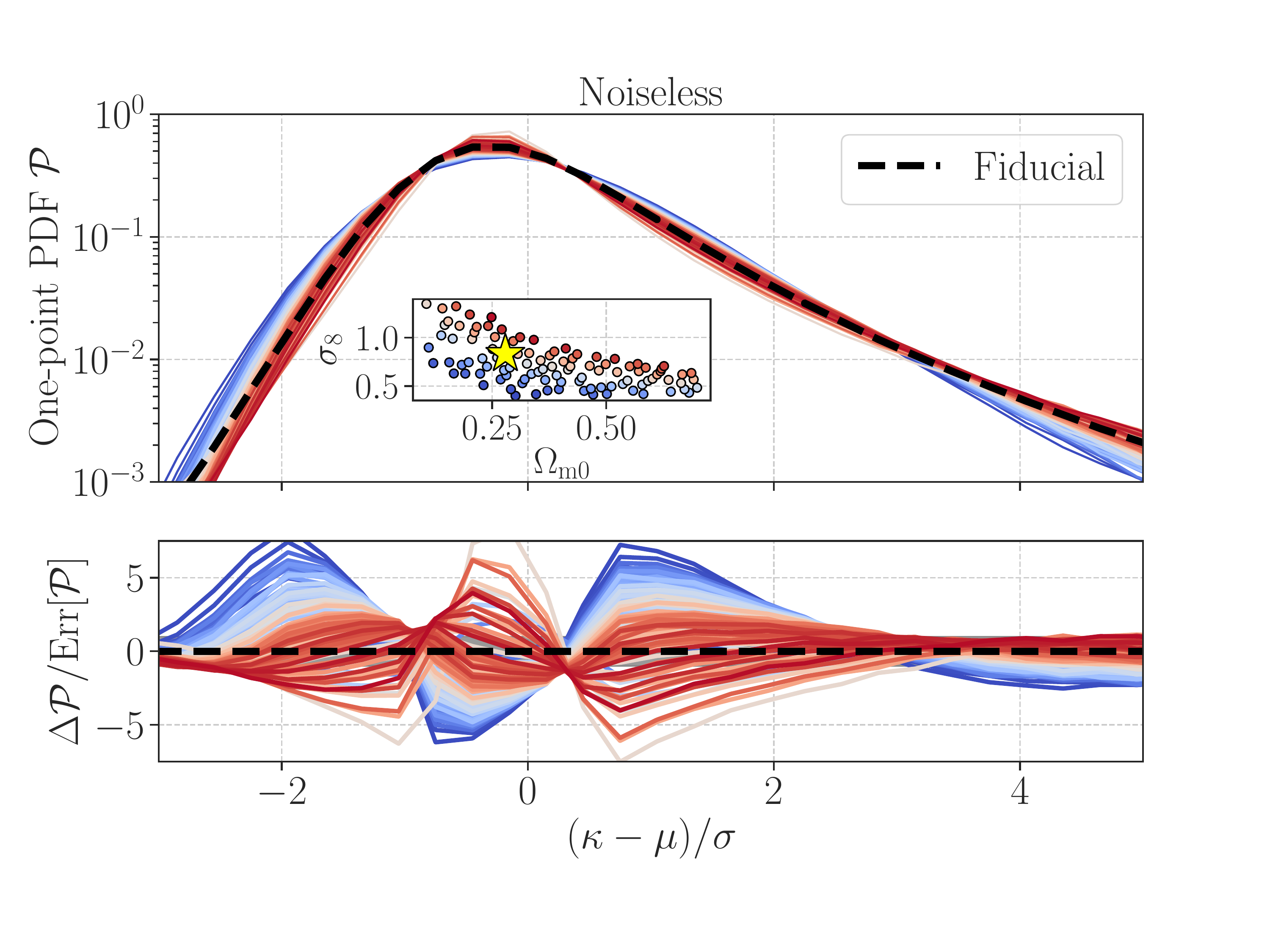}
     \caption{
     \label{fig:1pdf_true_cosmo}
     Similar to Figure~\ref{fig:1pdf_denoised_cosmo}, but for the lensing PDF in the absence of shape noises.
  } 
    \end{center}
\end{figure}

Figure~\ref{fig:1pdf_denoised_cosmo} summarises 
the cosmological dependence on the denoised PDF
for the HSC S16A. We find a clear dependence of 
the cosmological model,
highlighting that our GANs do not overfit to the assumed cosmology in the training.
The results in Figure~\ref{fig:1pdf_denoised_cosmo} can be 
compared with the PDFs for noisy input maps.
The cosmological dependence of the lensing PDFs without our denoising is shown in Figure~\ref{fig:1pdf_noisy_cosmo}.
This illustrates that the cosmological dependence is weak compared to the statistical error of the PDF when one works on the original noisy lensing map.
Our results indicate that the denoised PDF would potentially constrain
the cosmological parameters tighter than the noisy counterpart does.
However, the denoised PDFs are less sensitive to the cosmological parameters than 
the noiseless counterparts (see Figure~\ref{fig:1pdf_true_cosmo}).
In addition, the noiseless PDF is found to be 
largely sensitive to the parameter of $\sigma_8 (\Omega_{\rm m0}/0.3)^{0.2-0.3}$,
while the denoised and noisy PDFs are mainly determined by $\sigma_8 (\Omega_{\rm m0}/0.3)^{0.5}$.
According to these differences, 
the cosmological information in denoised PDFs is not identical to that in true PDFs.
We need additional analyses to study the information content in the denoised PDFs and 
its relation with other statistics \citep[e.g.][]{2017MNRAS.465.1974S}.
We leave those for our future study.

\subsection{Accounting for systematic uncertainties}

We here discuss generalisation errors in the denoising based on our GANs.
To quantify the errors, we introduce a simple chi-squared statistic for denoised lensing PDFs:
\beqa
\Delta \chi^2 = \sum_{i,j}
\left[{\cal P}_{i}(\mathrm{test})-{\cal P}_{i}(\mathrm{fid})\right]\, \bd{C}^{-1}_{ij}\, 
\left[{\cal P}_{j}(\mathrm{test})-{\cal P}_{j}(\mathrm{fid})\right],
\label{eq:chisq}
\eeqa
where ${\cal P}_i(\mathrm{test})$ is the denoised PDF at the $i$-th bin under our fiducial cosmological model but including different systematic effects in galaxy shape measurements, ${\cal P}(\mathrm{fid})$ is the denoised PDF for our fiducial model, and $\bd{C}$ represents the covariance matrix for the denoised PDF.
We evaluate ${\cal P}(\mathrm{test})$ by the average over 100 realisations shown in Sections~\ref{subsubsec:mock_photoz}, \ref{subsubsec:mock_mbias},
and \ref{subsubsec:mock_sigma_mea}.
Similarly, we compute ${\cal P}(\mathrm{fid})$ by averaging 1000 test data set in our fiducial run and estimate $\bd{C}$ from 1000 test realisations in our fiducial run.

Before showing results, we caution the limitation of our approach.
All the analyses in this paper assume the baryonic effects on the cosmic mass density can be negligible.
\citet{Osato:2015lja, Castro:2017tbn} examined the baryonic effects 
on the lensing PDF with 
hydrodynamical simulations and found the most prominent effect would appear in high-$\sigma$ tails in the PDF.
This is because the baryonic effects such as cooling, star formation, and feedback from active galactic nuclei commonly play a critical role in high-mass-density environments in the universe.
Besides, we ignore possible correlations between the lensing shear and the shape noises. An example causing such correlations is the intrinsic alignment (IA) \citep{Troxel:2014dba}.
Although this IA effect can potentially cause the biased parameter estimation in future surveys \citep{Krause:2015jqa}, we expect it would be less important for our analysis because we do not employ clustering
analyses of galaxy shapes. 
According to the observational facts, the IA effect is expected to be more prominent for redder galaxies (e.g. see \citet{Hirata:2007np}).
Since redder galaxies preferentially reside in denser environments such as galaxy clusters, we would mitigate the impact of the IA effect on our analysis when removing the high-$\sigma$ information.
To take into account these effects, 
we decide to remove such high-density regions 
by setting the range of pixel values to be ${\cal P}\ge0.01$ in Eq.~(\ref{eq:chisq}).
This leaves the lensing PDF at $-2.1 <(\kappa-\mu)/\sigma < 3.3$ with the number of bins being 18.
In this setup, we would argue that there exist significant generalisation errors in our denoising when finding $\Delta \chi^2 \ge \sqrt{2\times18}=6$.

\subsubsection{Photometric redshifts}


With our GANs, we assume the source redshift estimation by the specific {\tt mlz} method, but other methods predict the different redshift distributions accordingly (see Figure~\ref{fig:pz}).
To assess the systematic uncertainty due to imperfect photo-$z$ estimates, 
we compute Eq.~(\ref{eq:chisq})
when setting the term of ${\cal P}(\mathrm{test})$ to be the averaged PDF 
over 100 realisations of the mock catalogues with different photo-$z$ information (Sec~\ref{subsubsec:mock_photoz}).
We find that the photo-$z$ estimate by different methods can induce the bias in the lensing PDFs with a $\simlt0.3\sigma$ level over a wide range of pixel values.
These differences introduce $\Delta \chi^2 = 0.290$ and $0.134$ for 
the Photo-$z$ run 1 and 2, respectively.

\subsubsection{Multiplicative bias}

Besides, we assume the multiplicative bias defined by Eq.~(\ref{eq:mbias_for_single_pop}) 
is perfectly calibrated, but it can be mis-estimated with a level of 0.01.
To test this systematic effect, we input the average PDF obtained from the mock catalogues 
as described in Sec~\ref{subsubsec:mock_mbias} when computing Eq.~(\ref{eq:chisq}).
We find that a 1\%-level error in the multiplicative bias can induce the errors with a level of $\Delta \chi^2=0.238$ and 0.294 for $\Delta m_{\rm b} = 0.01$ and $-0.01$, respectively.

\subsubsection{Imperfect knowledge of noise}

We assume that the noise distribution in our mock catalogues is the same 
as in the real data. However, the actual measurement error of galaxy shapes 
is subject to a 10\%-level uncertainty.
To test the potential effect of this error, we input the average PDF obtained from the mock catalogues as described in Sec~\ref{subsubsec:mock_sigma_mea} in Eq.~(\ref{eq:chisq}).
We find that a 10\%-level mis-estimation in the shape measurement error
can the errors with a level of $\Delta \chi^2=0.213$ and 0.272 for $\Delta \sigma_{\rm mea} = 0.10$ and $-0.10$, respectively.

\begin{figure*}
\begin{center}
       \includegraphics[clip, width=2.2\columnwidth, viewport= 80 30 1008 864]
       {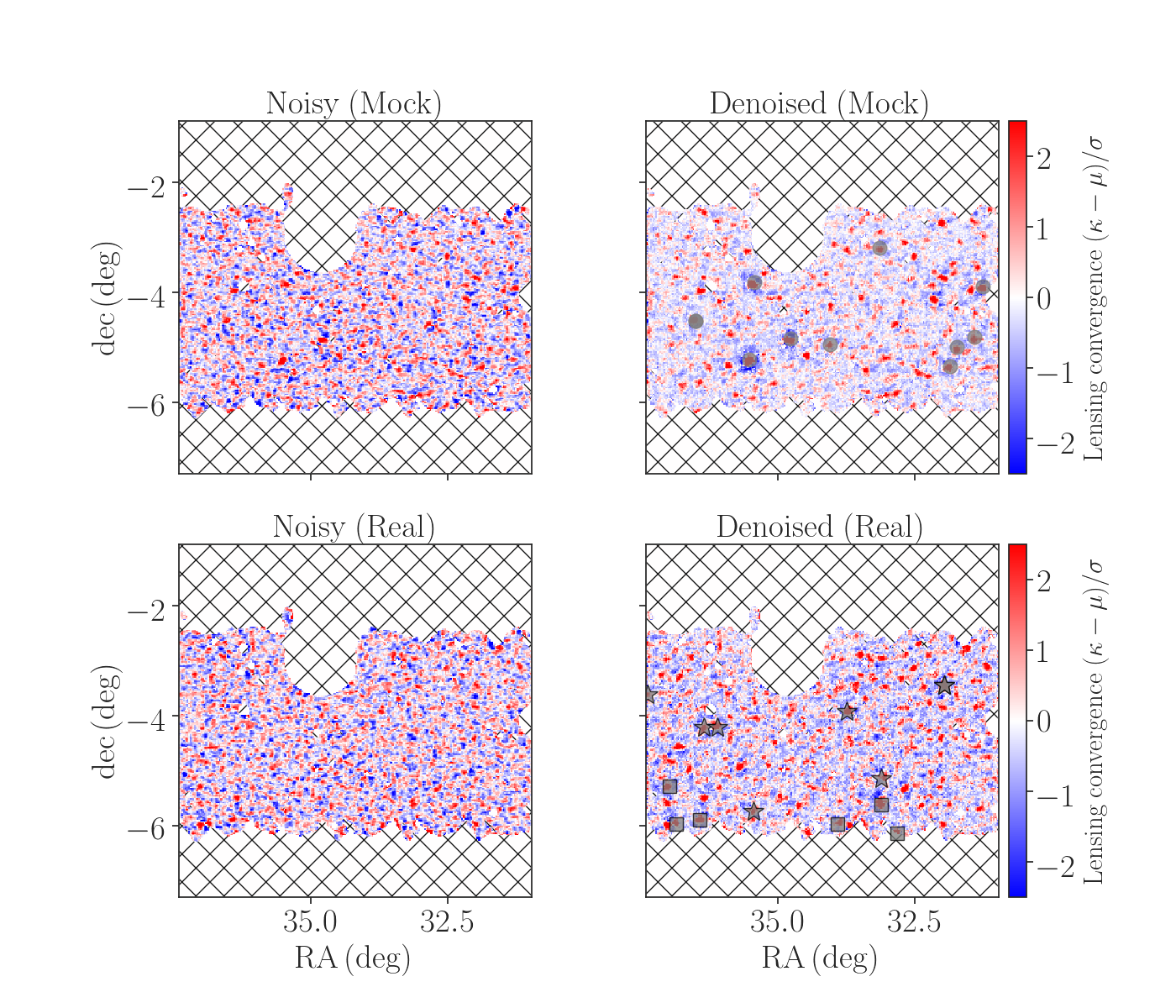}
     \caption{
     Performance of the denoising of the {\it observed} weak lensing map in the Subaru HSC first-year data.
     In upper panels, the left and right panels show a noisy input map and the denoised counterpart
     for a mock observation among 1000 realisations, respectively.
     The bottom panels show the results similar to the upper ones, but for the real observational data. In the top right panel, the grey points show the matched dark matter haloes to the local maxima on the denoised map. In the bottom left panel, the star and square symbols show 
     the matched galaxy clusters selected in optical and X-ray bands, respectively.
     Note that the hatched region represents the masked area because of missing the data.
     \label{fig:denoise_image_real}
  } 
    \end{center}
\end{figure*}

\subsubsection{Total systematic uncertainties}

Putting all together, we confirm $\Delta \chi^2 \simlt 1$ 
for the denoised PDFs. Hence, we conclude that 
possible systematic uncertainties in the measurement of galaxy shapes and redshifts are unimportant for the denoised PDF in our HSC data sets.
Now we are ready to apply our deep-learning denoising method 
to real HSC data.



\section{Application to Real data}
\label{sec:results}

\subsection{Visual investigation and cluster matching}

We apply our GAN-based denoising to the real weak lensing map obtained from the HSC S16A data.
Figure~\ref{fig:denoise_image_real} shows a comparison of the denoised images between mock and real data set.
The top left panel shows a noisy lensing field in a mock observation taken from 1000 realisations of the fiducial catalogues (Section~\ref{subsubsec:mock_fid}).
The top right panel represents the denoised weak lensing fields for the mock data.
In the bottom, the left and right panels are similar to the top, but for the real HSC S16A data.
On the denoised field, we mark the position of the matched galaxy cluster 
to the local maximum with its peak height greater than $5\sigma$.
For the mock data, we define the galaxy clusters by the dark matter haloes with their masses greater
than $10^{14}\, h^{-1}M_{\odot}$ and their redshifts $z<1$ in $N$-body simulations.
On the other hand, for the real data, we select the optically selected CAMIRA clusters \citep{Oguri:2017khw} in the HSC S16A with their richness of $>15$ and the X-ray selected clusters \citep{Adami:2018ysh} 
in our survey window by their X-ray temperature being $>2.14\, {\rm keV}$.
\citet{Oguri:2017khw} has shown that our selection of the optical richness and X-ray temperature
roughly corresponds to the selection of the cluster mass by $>10^{14}\, h^{-1}M_{\odot}$.
According to the results in Section~\ref{subsec:peak_halo_match}, we expect $\sim64.6\%$ of the peaks in a denoised field with their peak heights $>5\sigma$ will find their counterparts 
of galaxy clusters. In our denoised map for the real HSC S16A, we find 23 peaks and 13 peaks have the counterparts. This matching rate is in good agreement with the expectation from our experiments with 1000 mock observations. When limiting the CAMIRA clusters alone, we find 10 matched clusters in the denoised field, that is still consistent with our expectation within a $1\sigma$ Poisson error. Note that we have 3 matched clusters selected in both of optical and X-ray bands.

\subsection{Statistics-level comparisons}

\begin{figure}
\begin{center}
       \includegraphics[clip, width=1.1\columnwidth]
       {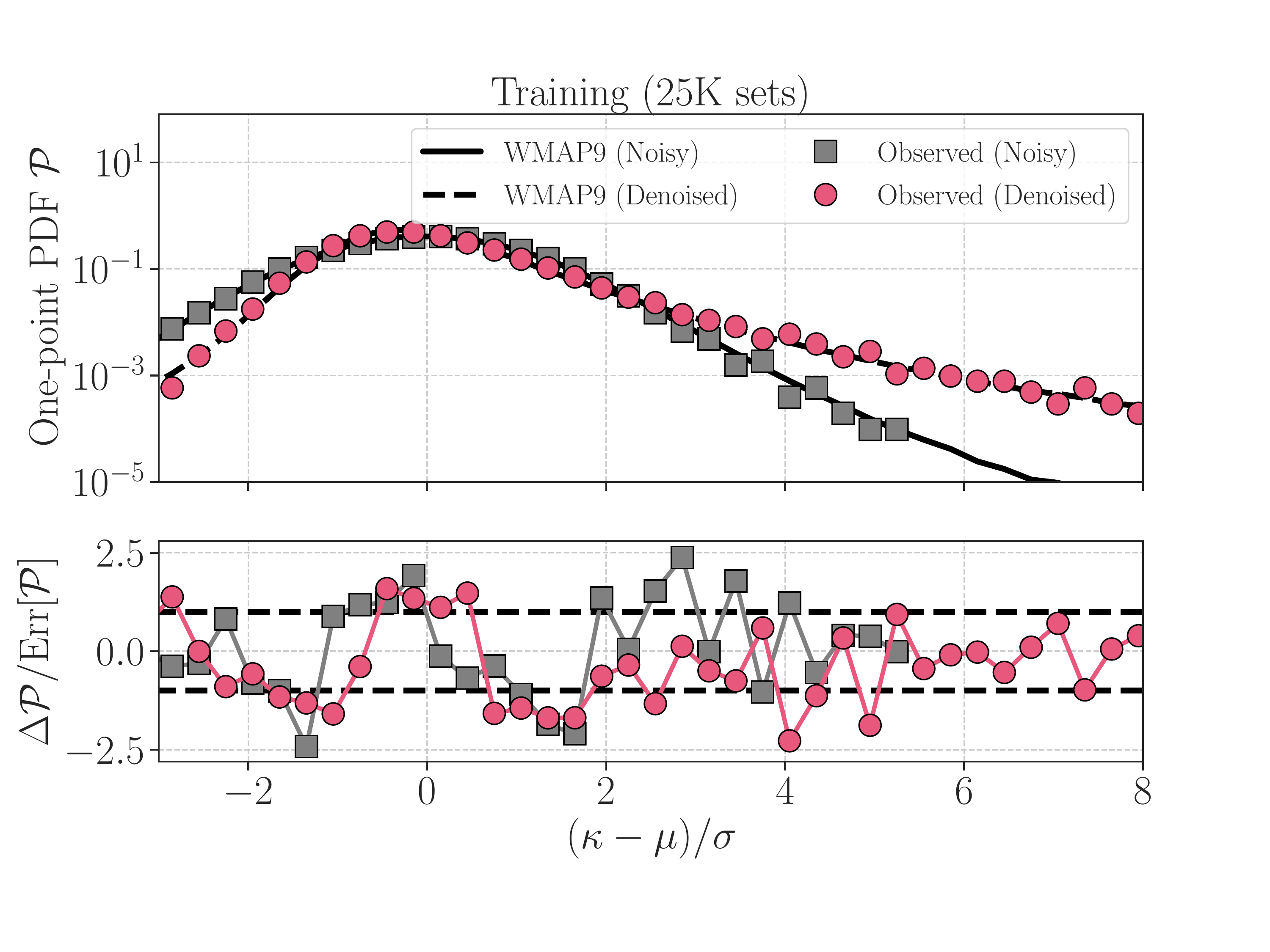}
     \caption{
     \label{fig:comp_WMAP9}
     The comparison of lensing PDFs between the HSC real data and WMAP9-cosmology prediciton. In the upper panel, the grey square and red circles show the observed PDF for noisy and denoised fields, respectively. We also present the corresponding predictions based on our mock observations under the WMAP9 cosmology with the lines.
     In the bottom, we show the difference between the observed PDF and the prediction in units of the sample variance. For a reference, the dashed lines show $\pm1\sigma$ levels.
  }
    \end{center}
\end{figure}

Finally, we employ a consistency test that our denoised field in the HSC is statistically consistent with our fiducial cosmology, i.e. WMAP9 cosmology \citep{Hinshaw:2012aka}.
Although we assume the WMAP9 cosmology in the training process, the denoised PDF by our GAN shows a cosmological dependence as shown in Figure~\ref{fig:1pdf_denoised_cosmo}.
Hence, it is not trivial if the denoised PDF for the real HSC data is consistent with the WMAP9 cosmology.
Figure~\ref{fig:comp_WMAP9} summarises the comparison of lensing PDFs between our measurement and the WMAP9-cosmology prediction.
For the consistency test, we introduce the following statistic:
\beqa
\chi^2 &=& 
\sum_{i,j}
\left[{\cal P}_{i}(\mathrm{obs})-{\cal P}_{i}(\mathrm{WMAP9})\right] \nonumber \\
&& \qquad
\times \, \bd{C}^{-1}_{ij}\, 
\left[{\cal P}_{j}(\mathrm{obs})-{\cal P}_{j}(\mathrm{WMAP9})\right],
\label{eq:chisq_obs}
\eeqa
where ${\cal P}(\mathrm{obs})$ is the observed PDF and 
${\cal P}(\mathrm{WMAP9})$ represents the prediction based on the WMAP9 cosmology.
We compute Eq.~(\ref{eq:chisq_obs}) for both of noisy and denoised PDFs.
We impose ${\cal P}(\mathrm{obs}) \ge 0.01$ in Eq.~(\ref{eq:chisq_obs})
to mitigate potential effects from baryons and IAs.
This setup remains 18 bins in the range of 
$-2.7 <(\kappa-\mu)/\sigma<2.7$ and 
$-2.1 <(\kappa-\mu)/\sigma<3.3$ for noisy and denoised PDFs, 
respectively.
We find $\chi^2=21.6$ and 16.3 for the noisy and denoised PDFs, respectively.
Hence, we conclude that our GAN-based denoised field 
is consistent with the WMAP9 cosmology.


\section{Conclusion and discussion}
\label{sec:con}

We have devised a novel technique for noise reduction of cosmic mass density maps 
obtained from weak lensing surveys.
We have improved over our previous analysis \citep{Shirasaki:2018thk} by 
incorporating realistic properties 
in the training data set
and by performing an ensemble learning with conditional GANs.
Our denoising method with GANs produces 10 estimates of the underlying noise field for a given input (observation). The multiple outputs allow us to reduce generalisation errors in the denoising process by taking the median value over 10 predictions by our networks.
For the first time, we have performed a stress-test on the denoising with deep-learning by using non-Gaussian statistics and by varying relevant parameters in mock lensing measurements.

Our findings through model validation are summarised as follows:

\begin{itemize}

\item Our GANs can reproduce the one-point probability distributions (PDFs) of noiseless fields with a level of $0.5-1\sigma$ level. This argument holds even when we vary 
the multiplicative bias with a level of 1\%, the photo-$z$ distribution of source galaxies, and the error in galaxy shape measurements with a level of 10\%.

\item After denoising, positive peaks with their pixel value greater than $5\sigma$ in lensing mass maps have counterparts of massive galaxies within a separation of 6 arcmins. The matching rate between peaks and clusters is found to be $\sim60\%$ for the denoised field, while it is comparable to the true matching rate of $\sim90\%$. We also confirmed that the matching in the denoised field is not a coincidence by studying the mass function of matched clusters.

\item Even though we assumed a specific cosmological model in the training of our GANs, the denoised field can show a cosmological dependence. The cosmological dependence on the denoised lensing PDF is different from the noiseless counterpart, whereas the sensitivity of the parameters $(\Omega_{\rm m0}, \sigma_8)$ in the denoised PDF is found to be greater than the noisy counterpart. This indicates that our GANs extract some cosmological information hidden by observational noises.

\end{itemize}

We have applied our method to the real observational data by using
a part of the Subaru Hyper-Suprime Cam (HSC) first-year shape catalogue \citep{2018PASJ...70S..25M}.
By comparing the denoised field for real HSC data with the prediction based on our mock observations, we concluded that our denoising provides a consistent result within the standard $\Lambda$CDM cosmological model.

The method developed in this paper can be easily generalised to other large-scale cosmological data sets \citep[e.g.][for intensity mapping surveys]{2020arXiv201000809M}.
Since the observable information is limited by the cosmic variance, 
future cosmology studies would need to extract information hidden behind 
observational noises within a limited data size.
Sophisticated modelling of cosmic large-scale structures can open a new window to produce mock observable ``universes'' as many as possible, which then allow us to redesign cosmological analyses beyond conventional methods.
Conditional GANs can provide an innovative approach for next-generation cosmological analyses. 
To gain full benefits from machine learning techniques in the future, we will need to solve technical problems of large-scale computing for deep learning and of fast and accurate modelling of the cosmic structure in multi-dimensional parameter space.
It would be necessary to develop methodologies to improve better understanding of neural networks in a physically-intuitive way.
A fruitful combination of astrophysics with machine learning is required to confront these challenges. 
Our results presented in this paper provide a 
prototype model for deep-learning-assisted cosmology, for further enhancement of the science returns in future astronomical surveys.

\section*{acknowledgments} 
This work was in part supported by Grant-in-Aid for Scientific Research on Innovative Areas from the MEXT KAKENHI Grant Number (18H04358, 19K14767), and by Japan Science and Technology Agency CREST Grant Number JPMJCR1414 and AIP Acceleration Research Grant Number JP20317829.
This work was also supported by JSPS KAKENHI Grant Numbers JP17K14273, and JP19H00677.
Numerical computations presented in this paper were in part carried out on the general-purpose PC farm at Center for Computational Astrophysics, CfCA, 
of National Astronomical Observatory of Japan.

The Hyper Suprime-Cam (HSC) collaboration includes the astronomical communities of Japan and Taiwan, and Princeton University. The HSC instrumentation and software were developed by the National Astronomical Observatory of Japan (NAOJ), the Kavli Institute for the Physics and Mathematics of the Universe (Kavli IPMU), the University of Tokyo, the High Energy Accelerator Research Organization (KEK), the Academia Sinica Institute for Astronomy and Astrophysics in Taiwan (ASIAA), and Princeton University. Funding was contributed by the FIRST program from Japanese Cabinet Office, the Ministry of Education, Culture, Sports, Science and Technology (MEXT), the Japan Society for the Promotion of Science (JSPS), Japan Science and Technology Agency (JST), the Toray Science Foundation, NAOJ, Kavli IPMU, KEK, ASIAA, and Princeton University. 

This paper makes use of software developed for the 
Vera C. Rubin Observatory.
We thank the LSST Project for making their code available as free software at \url{http://dm.lsst.org}.

The Pan-STARRS1 Surveys (PS1) have been made possible through contributions of the Institute for Astronomy, the University of Hawaii, the Pan-STARRS Project Office, the Max-Planck Society and its participating institutes, the Max Planck Institute for Astronomy, Heidelberg and the Max Planck Institute for Extraterrestrial Physics, Garching, The Johns Hopkins University, Durham University, the University of Edinburgh, Queen’s University Belfast, the Harvard-Smithsonian Center for Astrophysics, the Las Cumbres Observatory Global Telescope Network Incorporated, the National Central University of Taiwan, the Space Telescope Science Institute, the National Aeronautics and Space Administration under Grant No. NNX08AR22G issued through the Planetary Science Division of the NASA Science Mission Directorate, the National Science Foundation under Grant No. AST-1238877, the University of Maryland, and Eotvos Lorand University (ELTE) and the Los Alamos National Laboratory.

Based [in part] on data collected at the Subaru Telescope and retrieved from the HSC data archive system, which is operated by Subaru Telescope and Astronomy Data Center at National Astronomical Observatory of Japan.

\section*{Data Availability}
The data underlying this article will be shared on reasonable request to the corresponding author.

\appendix

\section{Additional validation tests of our GANs}\label{apdx:add_tests}
In this appendix, we show additional validation tests of our GAN performance.
The tests include two-point correlation analyses, a sanity check of over-fitting, 
and hyperparameter dependence on our results.
Note that we normalise a lensing map so that it has zero mean and unit variance in the tests.

\subsection{Clustering amplitudes}\label{apdx:2pcf}

\begin{figure}
\begin{center}
       \includegraphics[clip, width=1.1\columnwidth]
       {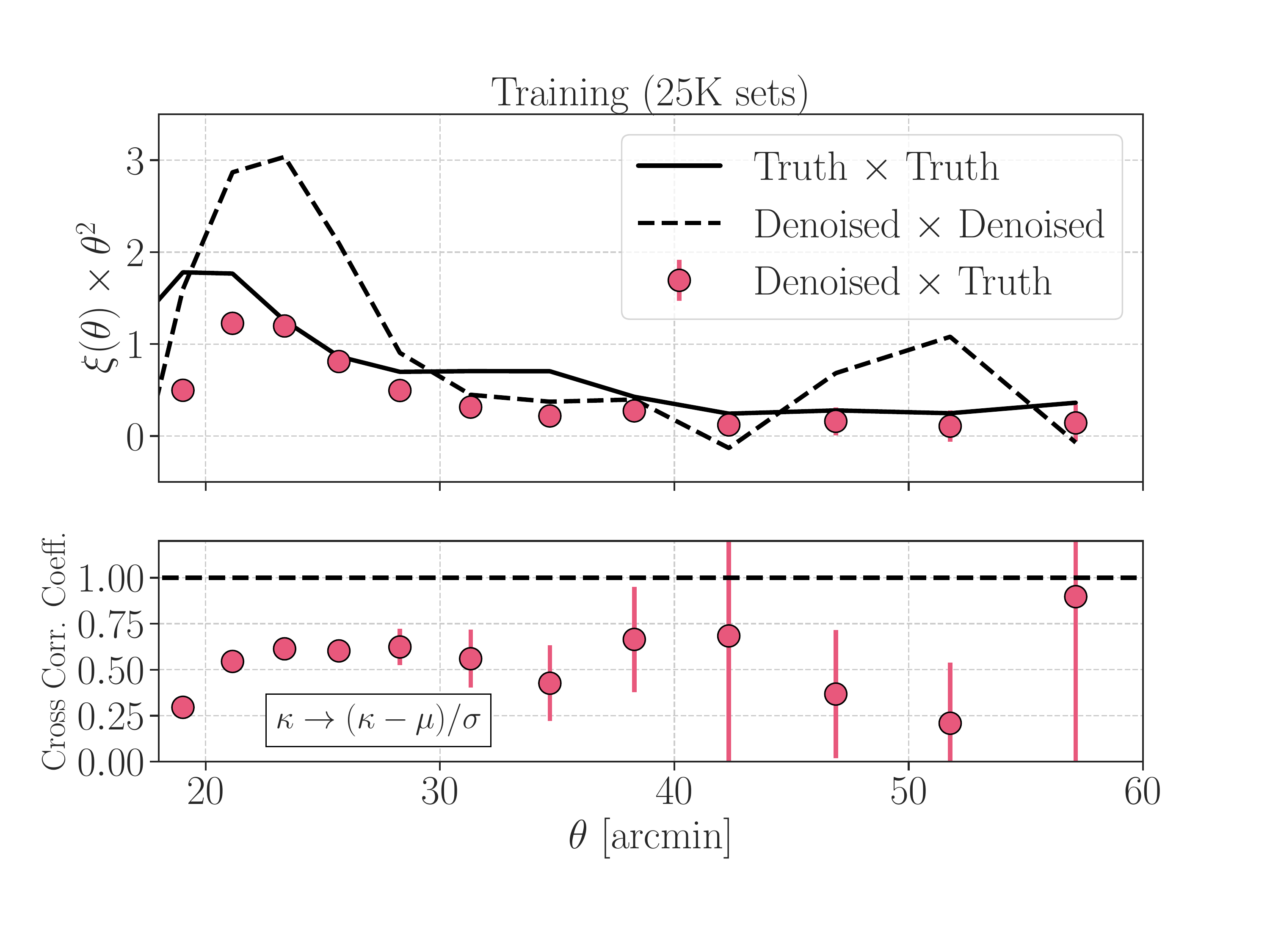}
     \caption{
     \label{fig:cross_corr}
     The two-point correlation analysis of noiseless and denoised fields.
     In the top panel, the points show the cross correlation between the noiseless and denoised fields,
     while the solid and dashed lines are for the 
     auto correlation of the noiseless and denoised fields.
     The bottom panel represents the cross-correlation coefficient in the two-point clustering.
  } 
    \end{center}
\end{figure}

To study the correlation between the denoised and 
the noiseless (true) fields, we perform a two-point correlation analysis.
For a given set of two random fields on a sky, we define the correlation function as
\beqa
\xi_{XY}(\theta) = \langle X({\bd \phi}) Y({\bd \phi}+{\bd \theta}) \rangle,
\eeqa
where $X$ and $Y$ are the two-dimensional random fields of interest.
We evaluate the two-point correlation functions for the noiseless and denoised fields
by using a public code {\tt TreeCorr} \citep{Jarvis:2003wq}.
We perform the linear-spaced binning 
in the angular separations from 3 to 60 arcmins,
but here focus on the large-scale clustering at 20-60 armins.
Figure~\ref{fig:cross_corr} shows 
the averages of the two-point correlation functions
over 1000 realisations. 

We find that the cross-correlation function between the noiseless and denoised fields is offset from the true auto correlation function, 
showing the denoised fields are biased for the ground truth.
Apart from the bias, the cross-correlation coefficient (CCC) is a measure of the correlation 
degree in the two-point correlation analysis. In our case, the CCC is defined as
\beqa
r = \frac{\xi_{{\rm true, denoised}}}{\sqrt{\xi_{{\rm true}}\, \xi_{{\rm denoised}}}}, \label{eq:ccc}
\eeqa
where $\xi_{{\rm true, denoised}}$ is the cross-correlation function between the noiseless and denoised fields,
while $\xi_{\rm true}$ is the auto correlation function of the noiseless field and so on.
The bottom panel in Figure~\ref{fig:cross_corr} shows that the CCC approaches $\simeq0.6$ on the large scales.
According to this, the denoised fields are found to surely correlate with the noiseless counterparts.
\rev{It would be worth noting that the auto correlation of noisy fields follows the noiseless counterpart, but the amplitude becomes smaller by $\sim0.25$ after the noisy field is normalised.}

\subsection{Statistical uncertainties in lensing PDFs}\label{apdx:stats_err_PDF}

\begin{figure}
\centering 
\includegraphics[clip,width=1.1\columnwidth]
{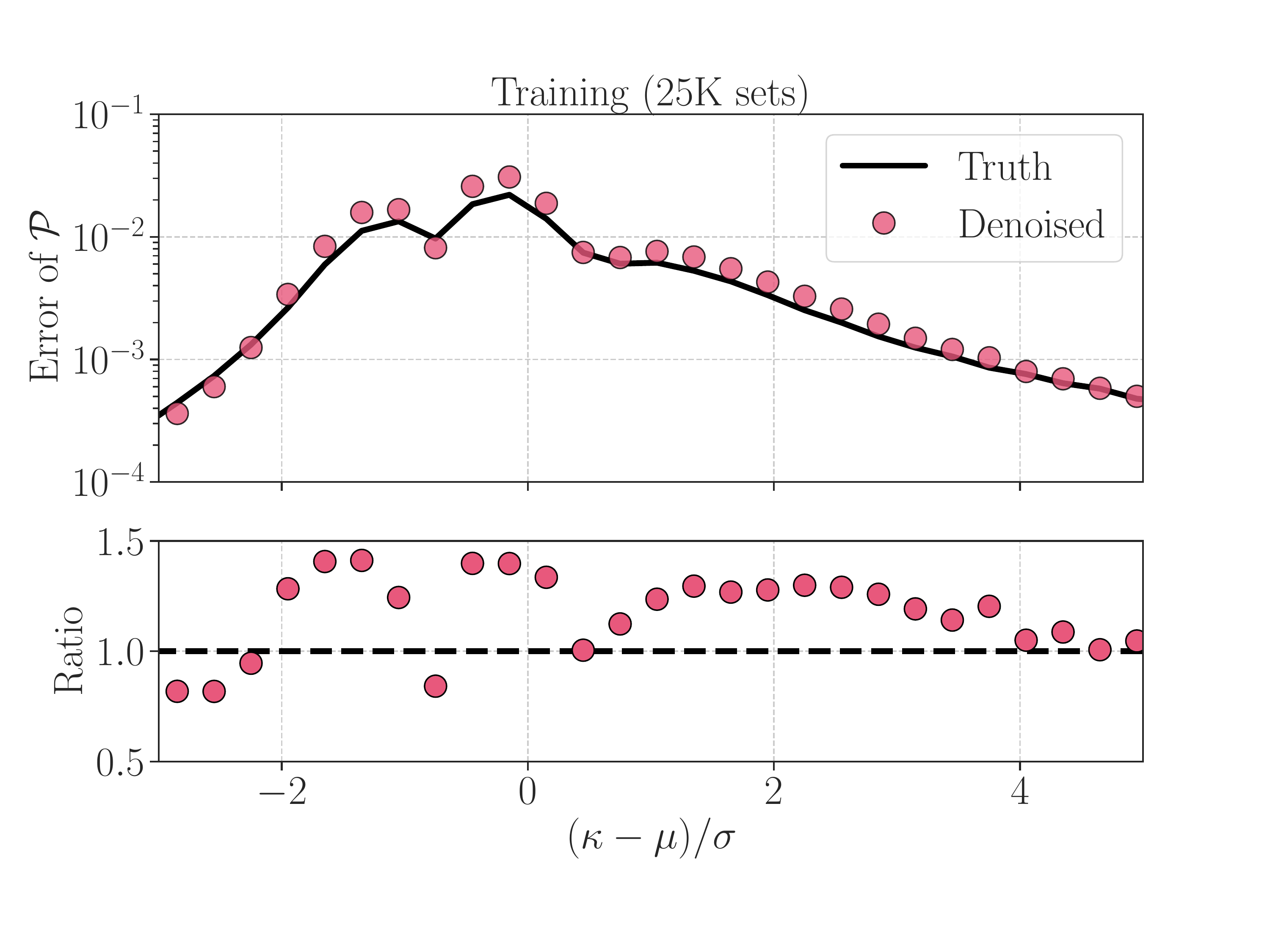}
\caption{
    The comparison of the variance in the one-point lensing PDF.
    In the upper panel, the solid line shows the variance for the true noiseless PDFs, while the points is the counterpart for the denoised PDFs. The ratio between two is also shown in the lower panel.
	\label{fig:denoise_pdf_err}
	}
\end{figure}

It is important to make sure that
our conditional GANs are not subject to 
over-fitting to our simulations.
In general, one can find over-fitting when the loss for test data sets 
becomes much smaller than the loss for training sets.
Because we are interested in statistical properties of the lensing map,
the comparison of losses is not always needed.
Instead, we compare the statistical uncertainty of 
the denoised PDF with that of the true counterpart.
We caution that our loss function of GANs is not designed to 
reconstruct noiseless lensing PDFs over realisations.
Hence, the lensing PDF after denoising should be regarded as the prediction by our GANs.
If the variance in the denoised PDFs becomes typically smaller than that of the true counterpart,
it makes the prediction by our GANs highly unreliable.
Figure~\ref{fig:denoise_pdf_err} shows the standard deviation of the lensing PDFs 
over the 1000 test data sets. 
The black line in the upper panel shows the true underlying scatter, 
while the red points are the denoised counterparts. 
We find that the statistical uncertainty in the denoised PDFs is larger 
than the intrinsic value in the wide range of lensing fields.  
This simple statistic indicates that the lensing PDFs by our GANs remain reliable.

\subsection{Varying a hyperparameter in the conditional GANs}\label{apdx:diff_lambda}

We here summarise the effect of a hyperparameter $\lambda$ in our deep-learning networks on the denoising performance.
To study the effect of $\lambda$, we consider additional two models with $\lambda=50$ and 100. We follow the same training strategy as in Section~\ref{sec:analysis} when varying $\lambda$.
We built 10 GANs for a given $\lambda$ and then estimated the ``best'' denoised field by using the median over 10 outputs by our GANs.

\begin{figure*}
\begin{center}
       \includegraphics[clip, width=2.2\columnwidth, viewport=100 50 1440 576]
       {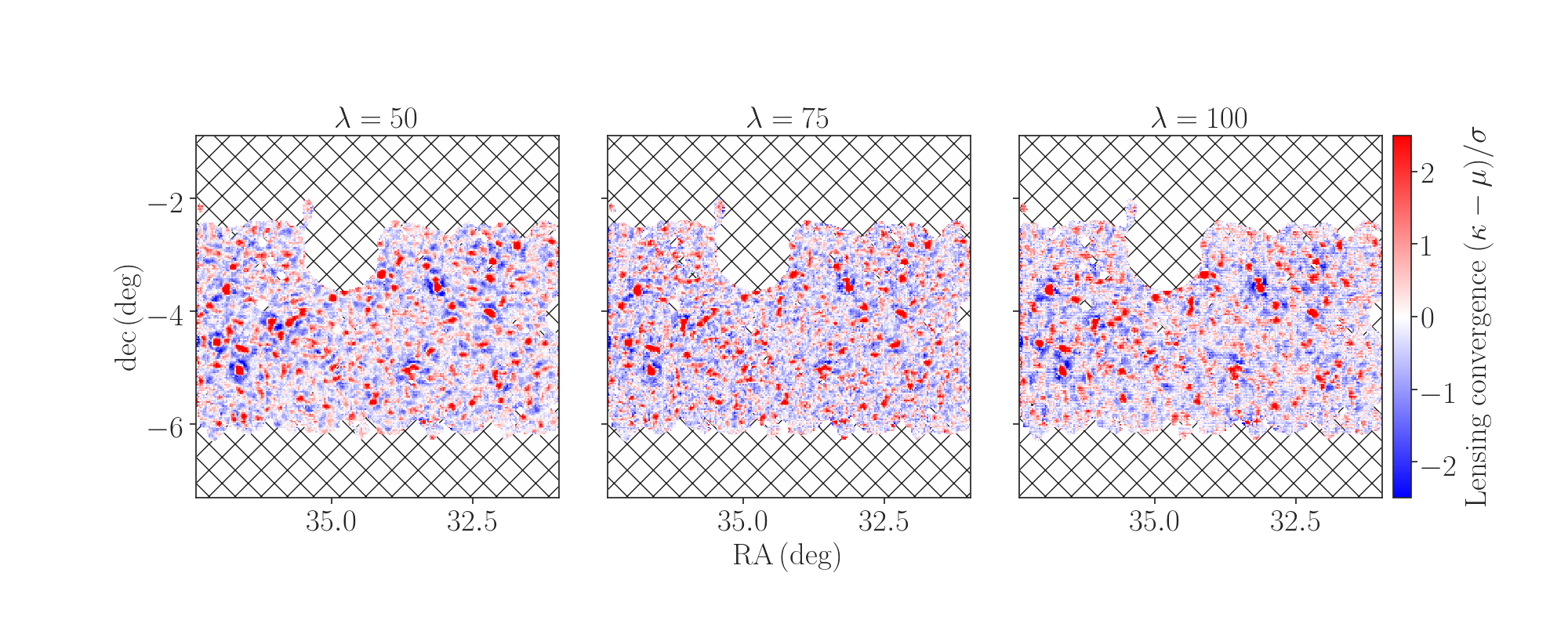}
     \caption{
     The effect of a hyperparameter $\lambda$ in our networks on image-to-image translation.
     In this figure, we work on the same realisation of lensing data in Figure~\ref{fig:train_image_fid}.
     From the left to right, we show the denoised lensing maps as varying the value of $\lambda = 50, 75$, and 100.
     \label{fig:train_image_diff_lambda}
  } 
    \end{center}
\end{figure*}

\begin{figure}
\begin{center}
       \includegraphics[clip, width=1.1\columnwidth]
       {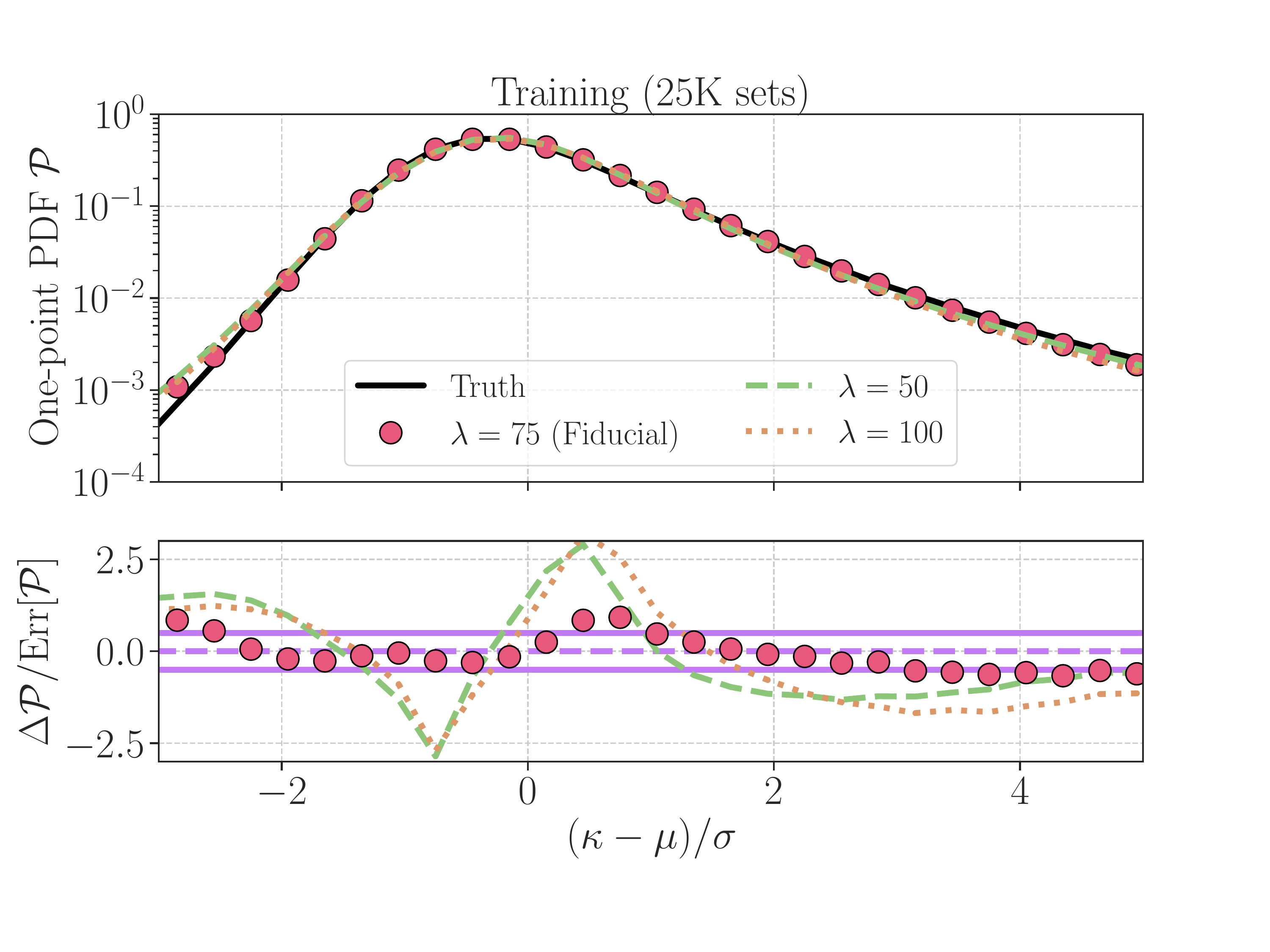}
     \caption{
     The effect of a hyperparameter $\lambda$ in our networks on denoised lensing PDFs.
     In the upper panel, the solid line shows the true noiseless lensing PDF, while the red circles represent the denoised counterpart with the default set up of $\lambda=75$.
     The green dashed and orange dotted lines stand for the denoised PDFs with $\lambda=50$ and 100, respectively.
     In the bottom, we show the difference between the denoised and noiseless PDFs in units of the sample variance in noiseless fields.
     For a reference, the magenta lines in the bottom show $\pm0.5\sigma$ levels.
     \label{fig:1pdf_diff_lambda}
  } 
    \end{center}
\end{figure}

\begin{figure}
\begin{center}
       \includegraphics[clip, width=1.1\columnwidth]
       {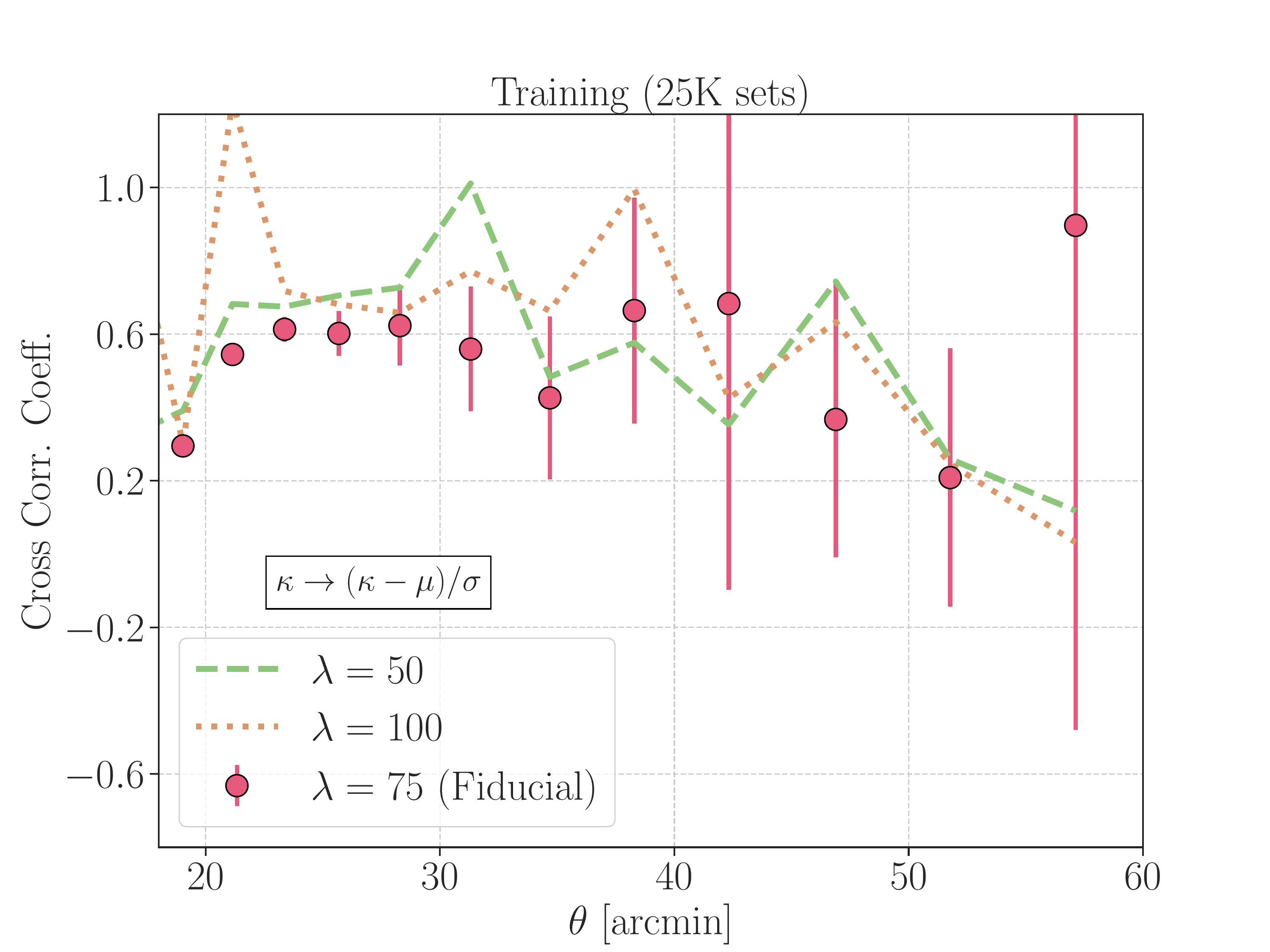}
     \caption{
     The effect of a hyperparameter $\lambda$ in our networks on the cross-correlation coefficient among the denoised, noisy, and noiseless fields. We define the coefficient in Eq.~(\ref{eq:ccc}). The red circles with error bars show our fiducial results with $\lambda=75$, while the green dashed and orange dotted lines represent the cases with $\lambda=50$ and 100, respectively.
     \label{fig:ccc_diff_lambda}
  } 
    \end{center}
\end{figure}

Figure~\ref{fig:train_image_diff_lambda} shows the comparison of the denoised map by GANs with different values of $\lambda$.
The figure highlights that the large-scale clustering pattern in the map looks less affected by the choice of $\lambda$, while the difference at over- and under-dense regions is prominent.
We also summarise the statistic-level comparisons in Figures~\ref{fig:1pdf_diff_lambda} and \ref{fig:ccc_diff_lambda},
supporting the visual implication.

\bibliographystyle{mnras}
\bibliography{refs}

\bsp	
\label{lastpage}

\end{document}